\documentclass[fleqn,usenatbib]{mnras}

\usepackage{txfonts}

\usepackage[T1]{fontenc}

\DeclareRobustCommand{\VAN}[3]{#2}
\let\VANthebibliography\thebibliography
\def\thebibliography{\DeclareRobustCommand{\VAN}[3]{##3}\VANthebibliography}

\usepackage{graphicx}	
\usepackage{natbib}
\usepackage{url}
\usepackage{longtable}
\usepackage{threeparttable}
\usepackage{multirow}
\usepackage{booktabs}
\usepackage{footnote}
\usepackage{wasysym}

\newcommand{\source}{MAXI~J1810$-$222}

\newcommand{\degree}{$^{\circ}$}

\newcommand{\arcsecond}{$^{\prime\prime}$}

\newcommand{\uJybeam}{$\mu$Jy\,beam$^{-1}$}

\newcommand{\swift}{\textit{Swift}}
\newcommand{\lrlx}{$L_{\mathrm{R}}$/$L_{\mathrm{X}}$}
\newcommand{\lEdd}{$L_{\mathrm{Edd}}$}

\newcommand{\lr}{$L_{\mathrm{R}}$}

\newcommand{\nh}{$N_{\mathrm{H}}$}

\newcommand{\nustar}{\textit{NuSTAR}}

\newcommand\T{\rule{0pt}{2.6ex}}       
\newcommand\B{\rule[-1.2ex]{0pt}{0pt}}

\graphicspath{{./}{Figures/}}

\title[Radio and X-ray observations of MAXI J1810$-$222]{Investigating the nature and properties of MAXI~J1810$-$222 with radio and X-ray observations}

\author[T. D. Russell et al.]{
T. D. Russell,$^{1}$\thanks{E-mail: thomas.russell@inaf.it}
M. Del Santo,$^{1}$
A. Marino,$^{2,1,3,4}$
A. Segreto,$^{1}$
S. E. Motta,$^{5}$
A. Bahramian,$^{6}$
\newauthor 
S. Corbel,$^{7,8}$
A. D'A{\`i},$^{1}$
T. Di Salvo,$^{1}$
J. C. A. Miller-Jones,$^{6}$
C. Pinto,$^{1}$
F. Pintore,$^{1}$
A. Tzioumis$^{9}$
\\
$^{1}$INAF, Istituto di Astrofisica Spaziale e Fisica Cosmica, Via U. La Malfa 153, I-90146 Palermo, Italy\\
$^{2}$Universit\`a degli Studi di Palermo, Dipartimento di Fisica e Chimica, via Archirafi 36 - 90123 Palermo, Italy\\
$^{3}$ Institute of Space Sciences (ICE, CSIC), Campus UAB, Carrer de Can Magrans s/n, E-08193 Barcelona, Spain \\
$^{4}$ Institut d'Estudis Espacials de Catalunya (IEEC), E-08034 Barcelona, Spain \\
$^{5}$INAF, Osservatorio Astronomico di Brera, via E. Bianchi 46, 23807 Merate (LC), Italy\\
$^{6}$ International Centre for Radio Astronomy Research -- Curtin University, GPO Box U1987, Perth, WA 6845, Australia\\
$^{7}$AIM, CEA, CNRS, Universit\'e de Paris, Universit\'e Paris-Saclay, F-91191 Gif-sur-Yvette, France\\
$^{8}$Station de Radioastronomie de Nan\c{c}ay, Observatoire de Paris, CNRS, PSL Research University, Univ. Orl\'eans, F-18330 Nan\c{c}ay, France\\
$^{9}$Australia Telescope National Facility, CSIRO, P.O. Box 76, Epping, New South Wales 1710, Australia}
\date{Accepted May 2022. Received May 2022; in original form May 2022}

\pubyear{2021}

\begin{document}
\label{firstpage}
\pagerange{\pageref{firstpage}--\pageref{lastpage}}
\maketitle

\begin{abstract}
We present results from radio and X-ray observations of the X-ray transient \source. The nature of the accretor in this source has not been identified. In this paper, we show results from a quasi-simultaneous radio and X-ray monitoring campaign taken with the Australia Telescope Compact Array (ATCA), the Neil Gehrels \swift\ observatory X-ray telescope (XRT), and the \swift\ Burst Alert Telescope (BAT). We also analyse the X-ray temporal behaviour using observations from the Neutron star Interior Composition Explorer (NICER). Results show a seemingly peculiar X-ray spectral evolution of \source\ during this outburst, where the source was initially only detected in the soft X-ray band for the early part of the outburst. Then, $\sim$200\,days after \source\ was first detected the hard X-ray emission increased and the source transitioned to a long-lived ($\sim 1.5$\,years) bright, harder X-ray state. After this hard state, \source\ returned back to a softer state, before fading and transitioning again to a harder state and then appearing to follow a more typical outburst decay. From the X-ray spectral and timing properties, and the source's radio behaviour, we argue that the results from this study are most consistent with \source\ being a relatively distant ($\gtrsim$6\,kpc) black hole X-ray binary. A sufficiently large distance to source can simply explain the seemingly odd outburst evolution that was observed, where only the brightest portion of the outburst was detectable by the all-sky X-ray telescopes.

\end{abstract}




\section{Introduction}
X-ray binaries (XRBs) consist of a compact object, either a stellar-mass black hole (BH) or a neutron star (NS), accreting matter from a companion star. Accretion typically occurs via Roche-lobe overflow, where the infalling, accreted matter forms a differentially rotating accretion disk as it spirals in toward the compact object \citep[e.g.,][]{1973A&A....29..179P}. However, not all of the infalling material is accreted onto the BH or NS primary star, instead some of it may be ejected from the system via outflows - either as a relativistic jet \citep[e.g.,][]{2006csxs.book..381F} or disk winds \citep[e.g.,][]{2016AN....337..368D}. While the accretion and wind phenomena are best observed in the optical and X-ray bands, the jets are observable at radio to infrared frequencies (and even up to X-rays and even Gamma rays). 

Transient XRBs spend the majority of their lifetimes in a low-luminosity quiescent state (with an X-ray luminosity, L$_{\rm X}$, of  $\lesssim 10^{33}$\,erg\,s$^{-1}$) but they may occasionally go into phases of outburst due to increased mass accretion onto the compact object (thought to possibly arise from disk instabilities; e.g., \citealt{1984AIPC..115...49V,2001NewAR..45..449L}) that can last weeks to years \citep[e.g.,][]{2016ApJS..222...15T}. During such outbursts their observable emission increases dramatically, where both the disk and jet emission increase by up to several orders of magnitude. Both BH and NS XRBs exhibit a variety of X-ray spectral states, broadly categorised as either hard or soft X-ray states. However, the outburst evolution for each source (or outburst) may vary \citep[e.g.,][]{2007A&ARv..15....1D}.

For BH XRBs, during a typical outburst the source is initially in a hard X-ray state, where the X-ray emission is dominated by a hard power-law ($\Gamma<2$) with a high-energy cut-off at $\sim$50-100\,keV. This component is usually interpreted as thermal Comptonisation of soft disk photons by a population of hot electrons ({\it{corona}}, $\sim$100\,keV) \citep{zdziarski04}. The hard state is associated with a flat to slightly inverted radio spectrum (with a radio spectral index $\alpha \gtrsim 0$, where the radio flux density, $S_{\nu}$, is proportional to the observing frequency, $\nu$, such that $S_{\nu} \propto \nu^{\alpha}$; e.g., \citealt{1979ApJ...232...34B}), arising from optically thick synchrotron emission originating from a persistent compact jet \citep[e.g.,][]{2001MNRAS.322...31F,2006csxs.book..381F,corbel02}. 

As the outburst progresses, the source brightens as the mass accretion rate increases, but remains in the hard X-ray state. At the same time the radio emission also brightens. At some point, the X-ray emission begins to soften as a thermal component that can be modelled with a multi-temperature black-body (with an inner disk temperature, kT$_{in}$, of $> 0.3$\,keV), starts to become increasingly important. This emission is thought to be due to an optically thick,  geometrically thin, accretion disk \citep{shakura73}.

As the disk luminosity (and temperature) increases and the X-ray emission softens. During this evolution the source can transition through the hard and soft intermediate states (HIMS and SIMS, respectively) on its way to the thermal dominant soft state \citep[e.g.,][]{delsanto08}. This progression is accompanied by changes in the X-ray spectral and timing properties \citep[e.g.,][]{belloni16}, as well as dramatic changes in the jet emission; during this progression, the compact jet switches off \citep{2004ApJ...617.1272C,2004MNRAS.355.1105F,2010LNP...794..115F,2020MNRAS.498.5772R}, being quenched by at least 3.5 orders of magnitude \citep{2011MNRAS.414..677C,2019ApJ...883..198R,2021MNRAS.504..444C} and a short-lived transient jet can be launched. The transient jet emission exhibits a steep radio spectrum ($\alpha \approx -0.6$) arising from ejected knots of (optically-thin) synchrotron emitting plasma \citep[e.g.,][]{2001MNRAS.322...31F}, possibly as they collide with the pre-existing jet or the surrounding environment \citep[e.g.,][]{2010MNRAS.401..394J,2017MNRAS.468.2788R}, although that is still debated. These discrete ejecta are launched close in time to the transition from the hard state to the soft state (the precise connection to the changing accretion flow is not well understood; e.g., \citealt{2009MNRAS.396.1370F,2012MNRAS.421..468M,2017MNRAS.469.3141T,2019ApJ...883..198R,2020NatAs.tmp....2B,2021MNRAS.505.3393W}). Remaining in the soft state, the source then typically begins to fade as the mass accretion rate decreases. During this state, jet emission is generally not detected although the discrete ejecta may be detected away from the object as they travel outwards \citep[e.g.,][]{2002Sci...298..196C,2019ApJ...883..198R,2020NatAs.tmp....2B,2021MNRAS.504..444C}.

As the source fades, the X-ray spectrum begins to harden once again, transitioning back to the hard state via the intermediate states in a reverse transition. Following its transition back to the hard state, the compact jet is progressively re-established over a period of several weeks \citep[e.g.,][]{2012MNRAS.421..468M,2013MNRAS.431L.107C,2013ApJ...779...95K,2014MNRAS.439.1390R}. The outburst ends when the source fades further, returning to its quiescent state, showing a fading hard X-ray spectrum as it does so.

Transient NS XRBs generally show a broadly similar pattern of behaviour during their outbursts \citep[e.g.,][]{2006MNRAS.366...79M, munoz2014}. However, the accretion-ejection picture is not as clear, possibly complicated by the presence of a stellar surface, boundary layer, and magnetosphere. The hard state for NS XRBs is associated with the launching of a compact jet, and these systems are capable of launching a transient jet during the hard-to-soft state transition \citep[e.g.,][]{2001ApJ...553L..27F,2004Natur.427..222F,2006MNRAS.366...79M}. However, while the compact jet has been observed to quench in the soft state of some systems \citep{2003MNRAS.342L..67M,2009MNRAS.400.2111T,2010ApJ...716L.109M,2017MNRAS.470.1871G,2020MNRAS.492.2858G}, it has not been identified in others \citep{1998ATel....8....1R,2003A&A...399..663K,2004MNRAS.351..186M,2011IAUS..275..233M}. The cause for this discrepancy is unclear. 

In their hard states, the radio ($L_{\rm R}$) and X-ray ($L_{\rm X}$) luminosities observed from BH XRBs appear to be coupled, showing a non-linear, empirical correlation between the luminosities \citep[e.g.,][]{1998A&A...337..460H,2000A&A...359..251C,2003A&A...400.1007C,2003MNRAS.344...60G,2012MNRAS.423..590G,2013MNRAS.428.2500C,2018MNRAS.478L.132G}. For NS systems, the \lrlx\ coupling is less straightforward, where individual systems can show various correlations, and not all sources follow a common or even defined track \citep[e.g.][]{2017MNRAS.470..324T,2021MNRAS.507.3899V}. These variations may be a result of having a NS primary (with a surface and magnetic field) impacting the observed emission from the accretion flow \citep[e.g.,][]{2008ASPC..401..191M}. It has also been suggested that NS XRB jets may show a different geometry or coupling to the accretion flow than in BH systems \citep{2020MNRAS.498.3351M}. NS XRBs are generally observed to be more radio faint than BH systems at a given X-ray luminosity \citep{2001MNRAS.324..923F,2006MNRAS.366...79M}, with the population typically being more radio faint by an average of $\sim$22 \citep{2018MNRAS.478L.132G}. Such a difference has been attributed to a number of different factors, such as jet power, primary mass, spin, magnetic field, and jet launching mechanism, although no direct connection has been identified yet \citep[e.g.,][]{2021MNRAS.507.3899V}. As such, although limited, a system's radio and X-ray luminosity has often been used to identify the nature of the accretor in XRBs. Although, we note that some NS systems can be as radio bright as their BH counterparts at similar X-ray luminosities \citep[e.g.,][]{2018ApJ...869L..16R,2020MNRAS.492.2858G}. NS XRBs can also be identified by the presence of thermonuclear Type-I X-ray bursts, which are sudden, short X-ray flashes that can be observed from accreting NS XRBs \citep{1993SSRv...62..223L}. Type-I bursts are thought to arise from the ignition and burning of accreted matter that has accumulated on the NS surface \citep[see][for review]{2021ASSL..461..209G}.

\subsection{MAXI~J1810$-$222}
\source\ was discovered on 2018 November 29 (MJD~58451; \citealt{2018ATel12254....1N}) by the Monitor of All-sky X-ray Image (MAXI) X-ray telescope on board the International Space Station \citep{2009PASJ...61..999M}, being identified as a soft X-ray transient. 
\nustar\ observation was performed on MJD~58461 (2018-12-09), showing a soft X-ray spectrum with no apparent X-ray pulsations or bursts \citep{2019ATel12398....1O}.
Due to Sun constraints, pointed X-ray observations did not occur again until the Neil Gehrels \textit{Swift} Observatory X-ray telescope (XRT) observed \source\ on 2019 February 09 ($\sim$MJD~58523). These first \swift-XRT observations measured the source position accurately and reported an X-ray spectrum that was still soft \citep{2019ATel12487....1K}. Analysis of UV and optical data from the \swift\ Ultra-Violet and Optical Telescope (UVOT) indicated a possible \textit{Gaia} Data Release 2\footnote{\citealt{2016A&A...595A...1G,2018A&A...616A...1G}} counterpart, implying a source distance\footnote{\citealt{2018AJ....156...58B}} of $730 \pm 30$\,pc \citep{2019ATel12487....1K}, although this would result in a low-luminosity soft state. 

Following the first \swift-XRT observation, \source\ faded steadily over the next $\sim$30\,days (until $\sim$MJD~58549). Early in this fading, when the source was still X-ray bright, radio observations taken on 2019 February 16 (MJD~58530) did not detect any radio emission from \source\ down to a 3-$\sigma$ upper-limit of 99\,\uJybeam\ \citep{2019ATel12521....1C}. Around MJD 58700 (2019 August 05), analysis of BAT survey data with the \texttt{BAT-IMAGER} software \citep{Segreto2010} showed an increase in the hard X-ray emission from this source, and we triggered a radio monitoring campaign with the Australia Telescope Compact Array. Over the next $\sim$few years \source\ was reported to brighten and harden on MJD 58667 (2019 July 03; \citealt{2019ATel12910....1N}), and again in 2020 late-February (from observations taken between MJDs~58906 and 58909; \citealt{2020ATel13540....1D}).

In this paper, we present radio and and X-ray monitoring of this source taken over a span of more than 2\,years to identify the nature of the accretor. In Section~\ref{sec:obs} we describe the details the radio and X-ray monitoring campaign. In Section~\ref{sec:results}, we outline the findings from these observations. In Section~\ref{sec:discussion}, we discuss the results, focusing on the nature of the compact object, its seemingly peculiar outburst evolution, and the distance to the source. We summarise our key findings in Section~\ref{sec:conclusions}.

\section{Observations \& Data Reduction}
\label{sec:obs}
\subsection{ATCA radio observations}

\source\ was observed by the Australia Telescope Compact Array (ATCA) fourteen times between late-2019 and late-2021 (see Tab. \ref{tab:ATCA_int}). \source\ was first observed by ATCA on 2019 November 13 (MJD~58800) but not again until 2020 December, following which higher cadence radio observations occurred between 2020 December and 2021 November. ATCA was in various configurations for these observations, but was mostly in its more extended 6\,km, 1.5\,km, and 750\,m configurations\footnote{\url{https://www.narrabri.atnf.csiro.au/operations/array_configurations/configurations.html}}. In all cases, the fixed location antenna 6 (located 6\,km from the array core) was used during the analysis, providing angular resolutions of $\sim$a few arcseconds for all observations. Most observations were recorded simultaneously at central frequencies of 5.5 and 9\,GHz, with a bandwidth of 2\,GHz comprised of 2048 1-MHz channels at each band. However, observations taken on 2021-05-17 and 2021-05-19 were composed of 32 1-MHz channels spaced evenly (every 64\,MHz) over 2\,GHz of bandwidth as well as 2048 finer (31.25\,kHz) channels within 64\,MHz of bandwidth for both frequency bands.

For all observations, the flux and bandpass calibration was carried out using PKS~1934$-$638, while the nearby ($\sim$3.67\degree\ away) J1817$-$254 was used for phase calibration. The data were edited for instrumental issues (shadowing etc) and radio frequency interference (RFI), calibrated, and imaged following standard procedures\footnote{\url{https://casaguides.nrao.edu/index.php?title=Main_Page}} in the Common Astronomy Software Application (\textsc{casa} version 5.1.2; \citealt{2007ASPC..376..127M}). The calibrated data sets were imaged using the \textsc{casa} task \textsc{clean}. Imaging used a Briggs weighting scheme with a robust parameter of 0, balancing sensitivity and resolution.

The flux density, $S_{\nu}$, of the source was measured using the \textsc{casa} task \textsc{imfit}, where the flux density of the point source was determined by fitting an elliptical Gaussian with full width at half maximum (FWHM) set by the synthesised beam shape. Errors on the absolute flux density scale include systematic uncertainties of 2\% for the 5.5/9\,GHz ATCA data\footnote{\url{https://www.atnf.csiro.au/observers/memos/d96783~1.pdf}} \citep[e.g.,][]{2011MNRAS.415.1597M,2012MNRAS.422.1527M}, which were added in quadrature with the root mean square (RMS) of the image noise. The radio luminosity, \lr, was calculated by $L_{\rm R} = 4 \pi S_{\nu} \nu D^{2}$, where $\nu$ is the observing frequency and $D$ is the distance to the source.

From our radio monitoring, we measure a best position (at 9\,GHz) of the radio counterpart to \source\ at a Right Ascension (R.A.) and Declination (Dec) of:\\
\\
R.A. (J2000) = 18$^{\rm h}$12$^{\rm m}$39.76$^{\rm s}$ $\pm$ 0.11$^{\prime\prime}$\\
Dec (J2000) = $-$22\degree19$^\prime$24.92$^{\prime\prime}$ $\pm$ 0.26$^{\prime\prime}$,\\
\\
where the errors are determined from the systematic errors (which are a function of the target's distance from the phase calibrator) added in quadrature with the larger of either the theoretical error of beam centroiding (beam size/2$\times$SNR, which dominates the R.A. error) or the statistical error on the fit (which dominates the Declination error).

\begin{table}
  \begin{center}
    \caption{Results from our ATCA radio observation. Each frequency band has a bandwidth of 2\,GHz. Flux density errors include systematic uncertainties. Upper-limits on the flux density are reported as $3 \times$ the RMS above the source position. We also provide the radio spectral index, $\alpha$, defined as $S_{\rm \nu} \propto \nu^{\alpha}$.}
    \begin{tabular}{c c c c c}
    \hline
          Date & MJD & Central & Flux & $\alpha$ \\
           & & frequency & density &  \\
          
          & & (GHz) & ($\mu$Jy) & \\
      \hline
      2019-11-13 & 58800.29 $\pm$ 0.06 & 5.5 & 570 $\pm$ 20 & $0.1 \pm 0.2$ \\
       & & 9.0 & 576 $\pm$ 16 &  \\
      2020-12-13 & 59196.08 $\pm$ 0.04 & 5.5 & 66 $\pm$ 17 & $0.0 \pm 0.5$ \\
       & & 9.0 & 65 $\pm$ 15 &  \\

      2020-12-27 & 59210.11 $\pm$ 0.09 & 5.5 & 73 $\pm$ 13 & $0.0 \pm 0.5$ \\
       & & 9.0 & 73 $\pm$ 12 &  \\

      2021-01-05 & 59219.20 $\pm$ 0.09 & 5.5 & 62 $\pm$ 15 & $-0.2 \pm 0.5$ \\
       & & 9.0 & 56 $\pm$ 15 &  \\

      2021-02-18 & 59263.80 $\pm$ 0.06 & 5.5 & 70 $\pm$ 20 & $0.3 \pm 0.5$ \\
       & & 9.0 & 80 $\pm$ 15 &  \\

      2021-03-07 & 59280.03 $\pm$ 0.09 & 5.5 & 75 $\pm$ 15 & $-0.3 \pm 0.6$ \\
       & & 9.0 & 60 $\pm$ 20$^{\rm a}$ &  \\

      2021-04-18 & 59322.88 $\pm$ 0.04 & 5.5 & 202 $\pm$ 18 & $0.0 \pm 0.3$ \\
       & & 9.0 & 200 $\pm$ 16 &  \\

      2021-04-30 & 59334.70 $\pm$ 0.06 & 5.5 & 325 $\pm$ 30 & $-0.07 \pm 0.30$ \\
       & & 9.0 & 315 $\pm$ 20 &  \\

      2021-05-17 & 59351.90 $\pm$ 0.03 & 5.5 & 240 $\pm$ 30 & $0.14 \pm 0.30$ \\
       & & 9.0 & 256 $\pm$ 20 &  \\

     2021-05-19 & 59353.86 $\pm$ 0.04 & 5.5 & 265 $\pm$ 40 & $0.4 \pm 0.3$ \\
       & & 9.0 & 320 $\pm$ 35 &  \\

     2021-05-30 & 59365.84 $\pm$ 0.06 & 5.5 & 120 $\pm$ 30 & $1.3 \pm 0.7$ \\
       & & 9.0 & 220 $\pm$ 25 &  \\

     2021-09-21 & 59478.31 $\pm$ 0.05 & 5.5 & 208 $\pm$ 16 & $-1.1 \pm 0.5$ \\
       & & 9.0 & 125 $\pm$ 18 &  \\

     2021-10-16 & 59503.21 $\pm$ 0.08 & 5.5 & $<$75 & -- \\
       & & 9.0 & $<$70 &  \\

     2021-11-13 & 59531.10 $\pm$ 0.12 & 5.5 & 225 $\pm$ 22 & $-0.4 \pm 0.3$ \\
       & & 9.0 & 185 $\pm$ 18 &  \\

       \hline
    \end{tabular}
    \label{tab:ATCA_int}
  \end{center}
$^{\rm a}$ This 9\,GHz observation suffered from some phase decorrelation due to poor weather, which we estimate to be of order 20\% and has been accounted for in the error (added in quadrature with the image and systematic errors).\\
\end{table}

\subsection{\it Swift}

\subsubsection{XRT}

\swift-XRT carried out a monitoring campaign on \source\  starting from 2019 February 16 until November 2021, for a total of 41 visits (Table \ref{tab:xrt-obst}). The XRT observations were downloaded from the \textsc{HEASARC} public archive\footnote{\url{https://heasarc.gsfc.nasa.gov/docs/archive.html}} (source ID: 00011105) and processed with \textsc{xrtpipeline} included within the \textsc{HEASOFT} software package (v. 6.26.1). The latest version of CALDB available when these data were released\footnote{\texttt{caldb.indx20190910}} was employed. We verified that the \source\ count-rate in the  observations performed in window timing (WT) mode was below 100 cts/s, thus resulting in a negligible pile-up impact. Observations in photon counting (PC) mode were moderately affected by pile-up ($\gtrsim$0.5 cts s$^{-1}$). With \textsc{ds9} we used circles of 20 pixels\footnote{As suggested by the \swift-XRT guidelines, e.g., \url{https://www.swift.ac.uk/analysis/xrt/spectra.php}} (corresponding to $\sim$47\arcsecond) for the source and background regions, while when pile-up was an issue (in PC mode) annuli were used where the inner and outer radius were dependent on the count-rate (see \url{https://heasarc.gsfc.nasa.gov/lheasoft/ftools/headas/xrtgrblc.html for discussion}). These regions were centered on the measured XRT position of R.A. (J2000) = 18$^{\rm h}$12$^{\rm m}$39.66$^{\rm s}$,
Dec (J2000) = $-$22\degree19$^\prime$25.0$^{\prime\prime}$ (with an error radius of 1.8\arcsecond\ after Swift-UVOT enhancement, consistent with our measured radio position). We processed the light curves and spectra by running the task \textsc{xrtproducts}. We rebinned each spectrum with \textsc{grppha} in order to have at least 25 counts per bin, which allows the use of $\chi^2$ statistics. Due to the relatively high column density and brightness of the source, we excluded data below 0.8\,keV. To explore if short-timescale X-ray flares and bursts were present in the observations we carefully checked the light curve of each observation, searching for intra-observational variability.

\begin{table} 
  \begin{center}
    \caption{List of the XRT observations of the source used in this work. ObsIDs have been shortened for brevity, where *=00011050. The MJD refers to the start of the X-ray observation.}
    \begin{tabular}{c c c c c }
    \hline
    ObsID & \multicolumn{2}{c}{Date} & {Obs. Mode} & {Exposure}   \\
    &  {(UTC)} & {(MJD)} & & {(ks)}  \\
     \hline
     *01 & 2019-02-09 & 58523.27 & PC & 0.96 \\
     *02 & 2019-02-16 & 58530.25 & WT & 0.98 \\
     *04 & 2019-02-19 & 58533.49 & WT & 0.99 \\ 
     *05 & 2019-02-21 & 58535.95 & WT & 0.96 \\
     *06 & 2019-02-23 & 58537.81 & WT & 0.99 \\
     *07 & 2019-02-24 & 58539.00 & WT & 0.88 \\
     *08 & 2019-03-03 & 58545.86 & WT & 0.96 \\
     *09 & 2019-03-05 & 58547.57 & WT & 0.98 \\
     *10 & 2019-03-07 & 58549.11 & WT & 1.06 \\
     *11 & 2019-03-09 & 58551.49 & WT & 0.92 \\ 
     *12 & 2019-03-11 & 58553.62 & WT & 0.38 \\
     *13 & 2019-03-13 & 58555.35 & WT & 0.87 \\ 
     *14 & 2019-07-05 & 58669.76 & WT & 0.99 \\
     *15 & 2019-11-05 & 58792.80 & PC & 0.99 \\
     *16 & 2021-02-09 & 59254.82 & PC & 1.00 \\
     *17 & 2021-02-18 & 59263.84 & PC & 0.86 \\
     *18 & 2021-02-25 & 59270.76 & PC & 0.93 \\
     *19 & 2021-03-10 & 59283.63 & PC & 0.95 \\
     *20 & 2021-03-28 & 59301.03 & PC & 0.89 \\
     *21 & 2021-04-11 & 59315.71 & WT & 0.55 \\
     *22 & 2021-04-15 & 59319.82 & PC & 0.99 \\
     *23 & 2021-04-18 & 59322.39 & PC & 0.83 \\
     *24 & 2021-04-21 & 59325.13 & PC & 1.00 \\
     *25 & 2021-04-25 & 59329.96 & PC & 0.63 \\
     *26 & 2021-04-25 & 59331.95 & WT & 0.42  \\
     *27 & 2021-05-03 & 59337.08 & WT & 0.10 \\
     *28 & 2021-05-09 & 59343.91 & WT & 0.97 \\
     *29 & 2021-05-12 & 59346.51 & PC & 0.03 \\
     *32 & 2021-05-23 & 59357.71 & PC & 0.09 \\
     *33 & 2021-05-26 & 59360.50 & PC & 0.88 \\
     *34 & 2021-05-30 & 59364.68 & PC & 0.92 \\
     *36 & 2021-06-06 & 59371.65 & PC & 0.90 \\
     *37 & 2021-08-20 & 59446.92 & PC & 0.18 \\
     *40 & 2021-09-08 & 59465.96 & PC & 0.89 \\
     *41 & 2021-09-16 & 59473.79 & WT & 0.20 \\
     *43 & 2021-09-21 & 59478.32 & PC & 0.86 \\
     *44 & 2021-09-30 & 59487.28 & PC & 0.97 \\
     *45 & 2021-10-06 & 59493.31 & PC & 0.91 \\
     *46 & 2021-10-13 & 59500.55 & PC & 1.01 \\
     *47 & 2021-10-20 & 59507.58 & PC & 1.36 \\
     *48 & 2021-10-27 & 59514.56 & PC & 0.87 \\
     *49 & 2021-11-03 & 59521.32 & PC & 1.44 \\
     \hline
     \end{tabular}
     \label{tab:xrt-obst}
  \end{center}
\end{table}

\begin{table} 
  \begin{center}
    \caption{List of the BAT observations of the source used in this work. The MJD ranges represent the time ranges from which survey BAT spectra were extracted. We also report the corresponding XRT obsID performed within the BAT time range, where the XRT obsIDs are given with the prefix of 00011050 removed (*=00011050).}
    \begin{tabular}{c c c c c c}
    \hline
          \multicolumn{6}{c}{BAT}\\
          \hline
    \multicolumn{2}{c}{Start Date} & \multicolumn{2}{c}{{End Date}} & {Exp.} & {XRT}  \\
  {(UTC)} & {(MJD)} & {(UTC)} & {(MJD)} & {(ks)}   & {Obs ID} \\
  \hline
   2019-06-06 & 58640.0 & 2019-08-03 & 58699.0 & 537 & *14 \\
   2019-10-04 & 58760.0 & 2020-01-01 & 58849.0 & 718 & *15 \\
   2021-01-10 & 59224.0 & 2021-03-10 & 59284.0 & 745 & *17 \\
   2021-03-27 & 59300.0 & 2021-04-26 & 59330.0 & 365 & *21 \\
   2021-04-26 & 59330.0 & 2021-05-16 & 59350.0 & 250 & *28 \\
   2021-05-30 & 59364.0 & 2021-07-13 & 59408.0 & 469 & *34 \\
   \hline
     \end{tabular}
     \label{tab:bat-obst}
  \end{center}
\end{table}

\subsubsection{BAT}
Data from the BAT survey were retrieved from the \textsc{HEASARC} public archive and processed using the \texttt{BAT-IMAGER} software \citep{Segreto2010}. This code processes data collected by coded mask telescopes. It computes all-sky maps after modelling and subtracting the proper background. Then, for each detected source, light curves and spectra can be extracted. 

We produced light curves of \source\ in count rate per pixel in the energy range 15 -- 50 keV, with 15 days binning time (Figure \ref{fig:lc}), as well as the 50 -- 80\,keV range. Spectra were extracted in the range 15 -- 90 keV, with logarithmic binning (for a total of 6 bins) and the official BAT spectral redistribution matrix was used. We checked to ensure that no other sources within the BAT field of view contaminated our results.

\subsection{\nustar}
\nustar\ observed \source\ on MJD~58461 (2018 December 9), with two pointings (ObsID 90410366001, 90410351001) of exposures of 0.8 and 12.7 ks, respectively. Among them, we analyzed only the observation with higher exposure. Data were reduced using the standard \textsc{Nustardas} task, incorporated in \textsc{Heasoft} (v. 6.26.1). The observation was taken in observing mode `06', which corresponds to the case when the on-board star tracker, the Camera Head Unit (CHU) \#4, is not available because it was either blinded by a bright source or by the Earth itself, so that the image reconstruction accuracy of  \nustar\ becomes profoundly degraded. The image of the source displays multiple centroids, so that an additional procedure is necessary to obtain reliable data. We used the task \textsc{nusplitsc} on cleaned event data in order to split the event files in several event files for each CHU combination (CHUs 3, 1+2 and 2+3 were used). After the treatment, we checked that the profile of the source was well defined and that it displayed a single centroid. For each event file, we extracted scientific products with the command \textsc{nuproducts} by using a circular region with a 60\arcsecond\ radius as the source region (centered at the radio position of the source), sufficient to encapsulate the source PSF. In order to take into account any background non-uniformity on the detector, for the background region we used four circular areas of 50\arcsecond\ in radius, placed on areas of the image with negligible contamination from the source. Finally, we summed the spectra corresponding to each CHU combination in a single spectrum. We repeated the whole procedure for both the two hard X-ray imaging telescopes on board \emph{NuSTAR}, i.e. the focal plane mirror (FPM) A and B. The last step was to re-bin the spectra, in order to have at least 25 counts per bin.

\subsection{NICER}

NICER observed \source\ 84 times between February 2019 to November 2020. A complete NICER spectral and timing analysis of \source\ is beyond the scope of this work (and will be presented in a forthcoming paper). As such, here we only selected three observations representative of the different spectral states we observed, in order to explore the overall variability properties of the source: we studied NICER observations with the identification numbers 1200560105, 2200560121, 2200560140. These observations were selected to be as close as possible to the XRT spectra with Obs IDs 0001105002, 00011105014, 00011105015 (see Section~\ref{sec:results}).

For each of the above observations we visually inspected the light curves and determined appropriate GTIs to remove instrumental artifacts. Then we extracted power density spectra (PDS) in the 0.5 -- 15.0 keV energy range in segments $\approx$15\,s long. We then averaged the Leahy-normalised PDS created from each segment to produce one averaged PDS per observation with a Nyquist frequency of $\approx$1000\,Hz. We did not correct for the contribution of the Poisson noise, but fitted it with a flat power law  when modeling the PDS.

\section{Results}
\label{sec:results}

\begin{figure*}
\centering
\includegraphics[width=1\textwidth]{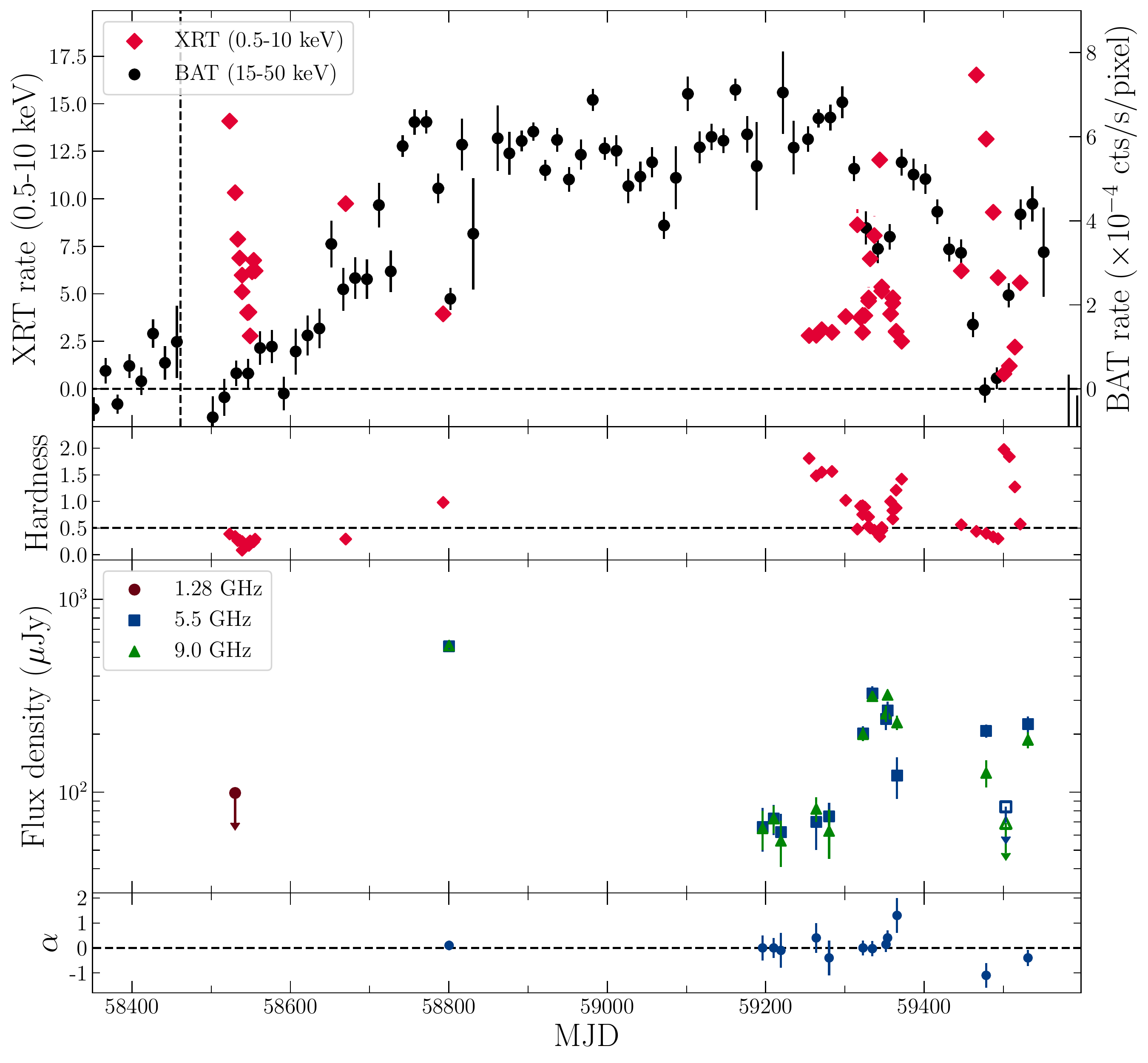}
\caption{X-ray and radio light curves of \source. \textit{First panel}: \swift-XRT (red diamonds) and BAT (black circles) X-ray light curves. The horizontal dashed line marks zero and the vertical dashed line represents the timing of the \nustar\ observation. \textit{Second panel}: X-ray hardness from the \swift-XRT monitoring, where hardness is defined as (2--10\,keV)/(0.5--2\,keV). As a guide, we arbitrarily place a horizontal line at a hardness of 0.5. \textit{Third panel}: Multi-frequency radio light curve, where the 5.5\,GHz and 9\,GHz data are shown as the blue squares and green triangles, respectively. The brown circle also shows the soft-state 1.28\,GHz non-detection reported by \citet{2019ATel12521....1C}, where we show the 3-$\sigma$ upper-limit. \textit{Fourth panel}: Radio spectral index, $\alpha$. The long-term X-ray and radio monitoring shows a seemingly atypical evolution during \source's long-lived outburst.}
\label{fig:lc}
\end{figure*}

\begin{figure*}
\centering
\includegraphics[width=0.9\textwidth]{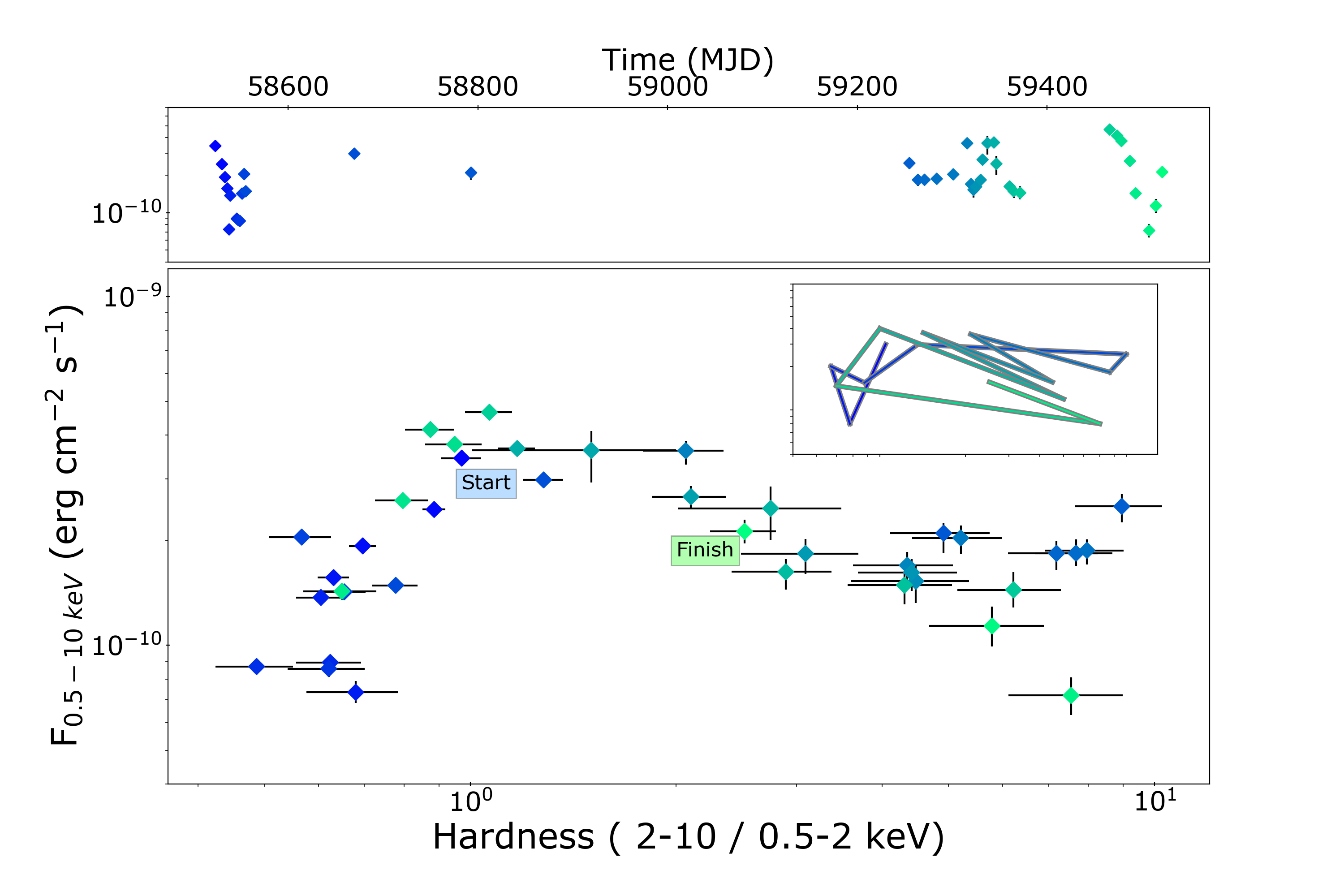}
\caption{XRT light curve (\textit{top panel}) and hardness intensity diagram (HID, \textit{lower panel}) of \source. The time sequencing of the data is highlighted with a colormap, from blue to green as shown in the top panel. The inset tracks out the hardness evolution during the outburst. Throughout its outburst, \source\ was observed to transition back and forth between bright softer and harder X-ray states, peculiar for a BH or NS XRB.}
\label{fig:hid}
\end{figure*}

In Figure \ref{fig:lc} we show the X-ray and radio light curves, X-ray hardness (defined as the count rate ratio 2--10 keV / 0.5 - 2\,keV) and radio spectral index, $\alpha$, obtained from our long-term monitoring campaign. The spectral evolution of the system is also shown by the XRT Hardness Intensity Diagram (HID; Figure \ref{fig:hid}). This HID has been computed by estimating the 0.5--2 keV / 2--10 keV fluxes with a spectral fitting procedure. Using \texttt{xspec}, we modelled each XRT spectrum with either a multi-color disk blackbody (\texttt{diskbb}) or a thermal Comptonization model (\texttt{nthcomp}), or a combination of the two models when required. In the spectral fits which included \texttt{nthcomp}, the \texttt{np\_type} parameter was set to 1, i.e. corresponding to disk blackbody seed photons, and the seed photons temperature tied to the $kT_{\rm disk}$ parameter when \texttt{diskbb} was included or fixed to 0.1 when no disk component was required. For all spectra, the interstellar absorption was left free and fitted with the model \texttt{tbabs}, using photoelectric cross sections from \cite{Verner1996} and element abundances from \cite{Wilms2000}. 
Spectral parameters of each XRT spectrum are shown in Table~\ref{tab:xrt_spectra}.

As apparent from Figure \ref{fig:hid}, the system displayed an atypical spectral evolution along the outburst. Based on the spectral variability, we selected a number of XRT spectra and combined them with quasi-simultaneous BAT spectra (Table~\ref{tab:bat-obst}). In the following sections, we report on the broad-band spectral fitting results as well radio evolution during the outburst.

\begin{table*}
\begin{center}
\caption{Fit results of the \emph{Swift}/XRT spectra. We used \texttt{diskbb}, \texttt{nthcomp} or both when required by the data. The \texttt{tbabs} model has been applied to all the spectra to take into account the neutral absorption.
The electron temperature of the Comptonisation component, $kT_{\rm e}$, has been fixed at 25 keV. 
Parameters in round parentheses were kept frozen. Quoted errors reflect 90\% confidence level. For the obsID *=00011050.}

\begin{tabular}{c  c  c c  c c  c }
\hline
Obs. &  $N_{\rm H}$ &  $kT_{\rm disk}$ & $K_{\rm disk}$ & $\Gamma$ & $F_{\rm 0.5-10 \ keV}$ & $\chi^2_\nu$ \T \B \\

&  ($\times$10$^{22}$ cm$^{-2}$)  & (keV) & &  &  ($\times 10^{-10}$ erg cm$^{-2}$ s$^{-1} $) & (d.o.f.) \T \B \\
\hline
*01 & $1.34^{+0.14}_{-0.12}$ & $0.53\pm{0.03}$ & $770^{+300}_{-210}$ & - & 3.44$^{+0.14}_{-0.11}$ & 0.96(76)  \T \B \\
*02 & $1.38\pm 0.07$ & $0.493^{+0.014}_{-0.013}$ & $830^{+160}_{-130}$ & - & 2.45$\pm$0.05  & 1.13(177) \T \B \\
*04 & $1.24^{+0.09}_{-0.08}$ &  $0.45\pm 0.02$ & $980^{+240}_{-190}$ & - & 1.93$\pm$0.04 & 0.76(150) \T \B \\
*05 & $1.21\pm0.08$ & $0.45^{+0.02}_{-0.01}$ & $840^{+210}_{-160}$ & - & 1.56$\pm$0.04 & 1.01(137) \T \B \\
*06 & $1.40\pm 0.20$ &   $0.42^{+0.04}_{-0.03}$ & $630^{+480}_{-260}$ & - & 0.73$^{+0.06}_{-0.04}$ & 0.91(36) \T \B \\
*07 & $1.27^{+0.10}_{-0.09}$ &  $0.43\pm 0.02$ &  $930^{+280}_{-210}$ & - & 1.37$\pm$0.04 & 1.01(124) \T \B \\
*08 & $1.27^{+0.13}_{-0.12}$ &  $0.41\pm 0.02$ &  $790^{+390}_{-240}$ & - & 0.89$^{+0.04}_{-0.03}$ & 1.06(93) \T \B \\
*09 & $1.14^{+0.12}_{-0.11}$ & $0.40\pm 0.02$  & $870^{+370}_{-250}$ & - & 0.87$^{+0.04}_{-0.03}$ & 0.98(99) \T \B \\
*10 & $1.46^{+0.16}_{-0.15}$ & $0.41\pm 0.02$   & $880^{+460}_{-290}$ & - & 0.86$^{+0.05}_{-0.04}$  & 0.94(72) \T \B \\
*11 & $1.15\pm0.09$ & $0.45\pm0.02$ & $700^{+210}_{-160}$ & - & 1.42$\pm$0.04 & 0.96(131) \T \B \\   
*12 & $1.22^{+0.15}_{-0.13}$ & $0.43\pm0.02$ &  $1360^{+670}_{-420}$ & - & 2.04$^{+0.08}_{-0.07}$  & 1.07(58) \T \B \\ 
*13 & $1.43^{+0.11}_{-0.10}$ & $0.45\pm0.02$ & $850^{+270}_{-200}$  & - & 1.48$\pm$0.05 & 0.87(127) \T \B \\ 
*14 & $0.74\pm 0.05 $  & $0.38\pm{0.02}$ & 1800$^{+920}_{-570}$ & $2.1^{+0.3}_{-0.1}$ & 2.98$^{+0.11}_{-0.10}$ &  0.91(200) \T \B \\
*15 & 0.4$\pm$0.01 & - & - & $1.6 \pm 0.1$ & 2.10$^{+0.20}_{-0.15}$ & 1.17(55) \T \B \\
*16 & $1.0\pm0.3$ & - & - & 1.38$^{+0.19}_{-0.17}$ & 2.50$\pm$0.20 & 0.93(42) \T \B \\
*17 & $1.0\pm0.2$ & - & - & 1.63$^{+0.07}_{-0.06}$ & 1.80$\pm$0.20 & 0.77(45) \T \B \\
*18 & $1.3\pm0.3$ & - & -  & $1.7\pm0.2$ & 1.80$^{+0.15}_{-0.17}$  & 0.95(40) \T \B \\
*19 & $1.04^{+0.12}_{-0.10}$ & - & - & $1.47\pm0.05$ & 1.87$^{+0.19}_{-0.16}$ & 0.78(41) \T \B \\
*20 & $0.70\pm 0.20$ & - & - & $1.58\pm 0.20$ & 2.00$\pm$0.20 & 1.10(38) \T \B \\
*21 & $1.08^{+0.15}_{-0.12}$ & $0.32\pm0.02$ & 3825 $^{+4230}_{-2250}$ & 1.8$^{+0.1}_{-0.2}$ & 3.60$^{+0.30}_{-0.20}$ & 0.92(131) \T \B \\
*22 & $0.9\pm0.2$ & - & - & $1.9\pm0.2$ & 1.70$^{+0.19}_{-0.16}$ & 0.78(28) \T \B \\
*23 & $0.6^{+0.3}_{-0.2}$ & - & - & $1.6\pm0.2$ & 1.50$\pm$0.20 & 0.81(21) \T \B \\
*24 & $1.0\pm0.3$ & - & - & $2.0\pm0.2$ & 1.62$^{+0.18}_{-0.15}$ & 0.78(28) \T \B \\
*25 & $1.0\pm0.3$ & - & - & $2.2\pm 0.3$ & 1.80$\pm$0.20 & 0.78(28) \T \B \\
*26 & $1.0\pm0.2$ & $0.31^{+0.08}_{-0.07}$ &  $2700^{+1200}_{-1900}$ &  $1.8\pm0.4$ & 2.70$\pm$0.20 & 0.90(81) \T \B \\
*27 & $0.90^{+0.40}_{-0.30}$  & - & - & $2.8\pm0.5$ & 3.60$^{+0.60}_{-0.40}$ & 0.90(14) \T \B \\
*28 &  $1.13\pm0.08$ & $0.35\pm0.03$ & 3895$^{+2330}_{-1330}$ & 2.1$\pm0.3$ & 3.67$^{+0.12}_{-0.11}$ & 1.02(211) \T \B \\
*29 & (1.0) &  $0.35^{+0.07}_{-0.06}$  & $1310^{+1600}_{-710}$ & <2.1 & 2.50$^{+0.50}_{-0.30}$ & 0.90(22) \T \B \\
*33 & $0.9\pm0.2$ & - & - & $2.2\pm0.2$ & 1.62$^{+0.18}_{-0.14}$ & 1.00(29) \T \B \\
*34 & $0.8\pm0.2$ & - & - & $1.8\pm0.2$ & 1.49$^{+0.18}_{-0.14}$ & 0.89(24) \T \B \\
*36 & $0.7\pm0.2$ & - & - & $1.7^{+0.3}_{-0.2}$ & 1.44$^{+0.16}_{-0.18}$ & 0.98(20) \T \B \\
*40 & $1.43^{+0.16}_{-0.15}$ & 0.54$\pm$0.03 & $990^{+430}_{-290}$ & - & 4.65$^{+0.18}_{-0.16}$ & 1.21(71) \T \B \\
*41 & 1.02$\pm$0.11 & 0.55$\pm$0.03 & 660$^{+240}_{-170}$ & - & 4.15$\pm$0.16 & 0.99(72) \T \B \\ 
*43 & $1.41^{+0.19}_{-0.16}$ & 0.52$\pm$0.04 & 1040$^{+570}_{-350}$ & - & 3.77$^{+0.18}_{-0.17}$ & 1.20(55) \T \B \\
*44 & $1.31^{+0.16}_{-0.15}$ & 0.49$\pm$0.03 & 920$^{+440}_{-280}$ & - & 2.59$\pm$0.11 & 0.87(56) \T \B \\
*45 & $1.35^{+0.20}_{-0.18}$ & 0.44$\pm$0.03 & 900$^{+660}_{-350}$ & - & 1.42$\pm$0.08  & 0.94(37) \T \B \\
*47 & (1.40) & - & - & $1.7\pm 0.2$ & 0.72$^{+0.09}_{-0.08}$  & 0.78(13) \T \B \\
*48 & 1.20$\pm$0.50 & - & - & $1.9\pm 0.3$ & 1.10$^{+0.14}_{-0.15}$  & 0.95(15) \T \B \\
*49 & $0.80^{+0.30}_{-0.20}$ & $0.47^{+0.18}_{-0.15}$ & $170^{+870}_{-120}$ & $1.8^{+0.5}_{-0.8}$ & 2.12$^{+0.16}_{-0.15}$ & 1.3(57) \T \B \\

\hline
\end{tabular}
\label{tab:xrt_spectra}
\end{center}
\end{table*}

\subsection{Outburst evolution in the X-ray and radio}

\subsubsection{Initial discovery in the soft state}

\begin{figure}
\centering
\includegraphics[width=\columnwidth]{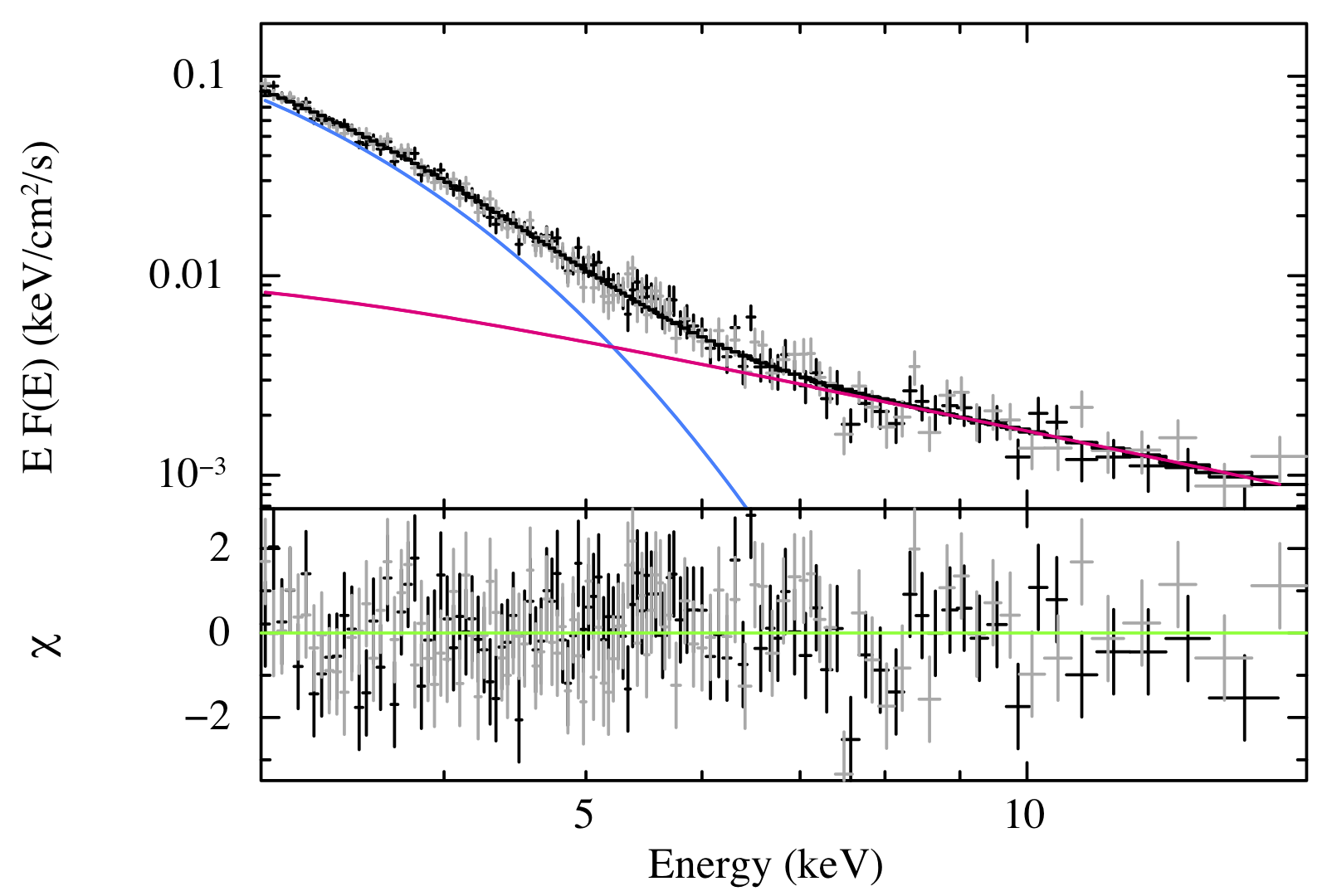}
\caption{Residuals for the {\it NuSTAR} observation performed during the soft state of the system. Data: FPMA (black) and FPMB (grey). The blue (magenta) thick line indicates the blackbody (\texttt{nthcomp}) contributions to the total spectrum. The \nustar\ spectrum was well described by a disk black body with a (steep) hard X-ray powerlaw component.}
\label{fig:nust-spectra}
\end{figure}

The discovery X-ray detection and X-ray follow-up soon after its initial detection implied a soft X-ray spectrum, as reported by both {\it MAXI} and {\it NuSTAR} teams \citep{2009PASJ...61..999M,2019ATel12398....1O}. Indeed, the {\it NuSTAR} spectrum (see Figure \ref{fig:nust-spectra}) can be well described by a $\sim$ 0.5 keV disk blackbody and a steep hard X-ray tail with $\Gamma \sim 3.5$ (see Table~\ref{tab:broadband} for details on the spectral fitting parameters). In addition, \source\ was not significantly detected with the BAT hard X-ray telescope (Figure~\ref{fig:lc}), confirming that the system was in a softer X-ray state. However, it is noteworthy that there is a hint of a potential brightening in stacked BAT observations from early 2018 November (from MJD~58426.6 $\pm$ 7.5) - in the weeks before the first reported X-ray detection of the source (see Figure \ref{fig:lc} showing the marginal rise).

\begin{figure*}
\centering
\includegraphics[width=\columnwidth]{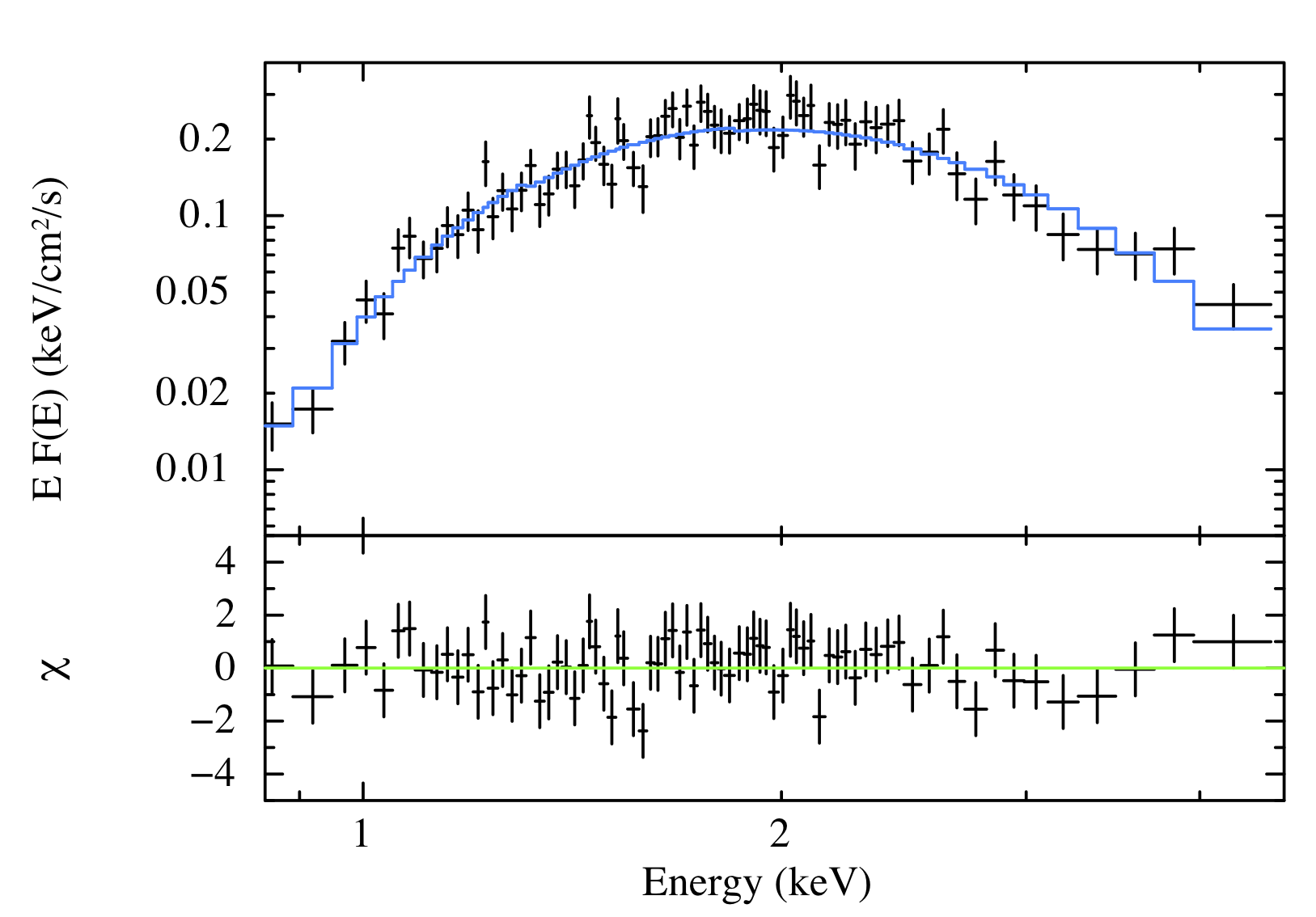}
\includegraphics[width=0.9\columnwidth]{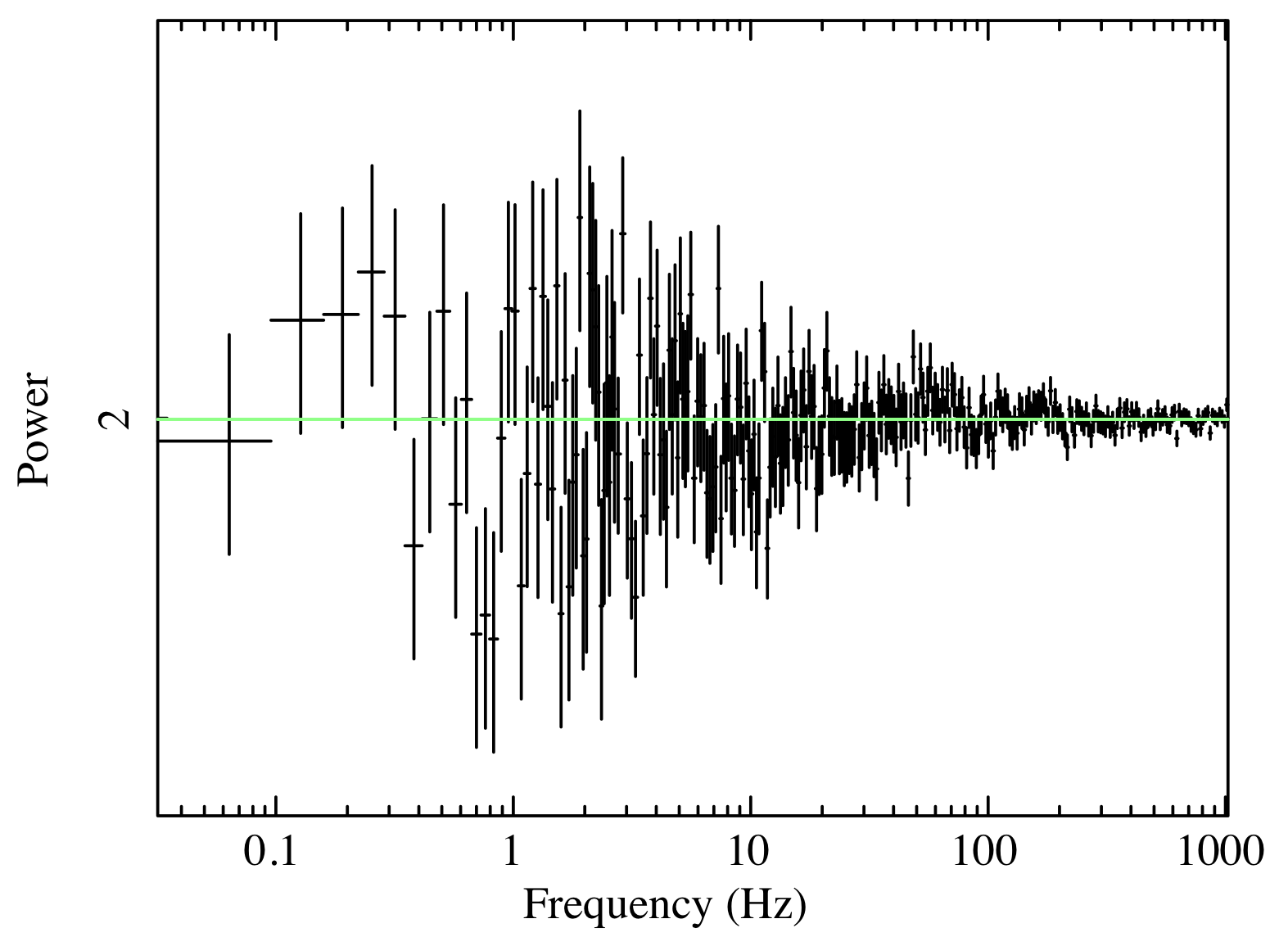}
\includegraphics[width=\columnwidth]{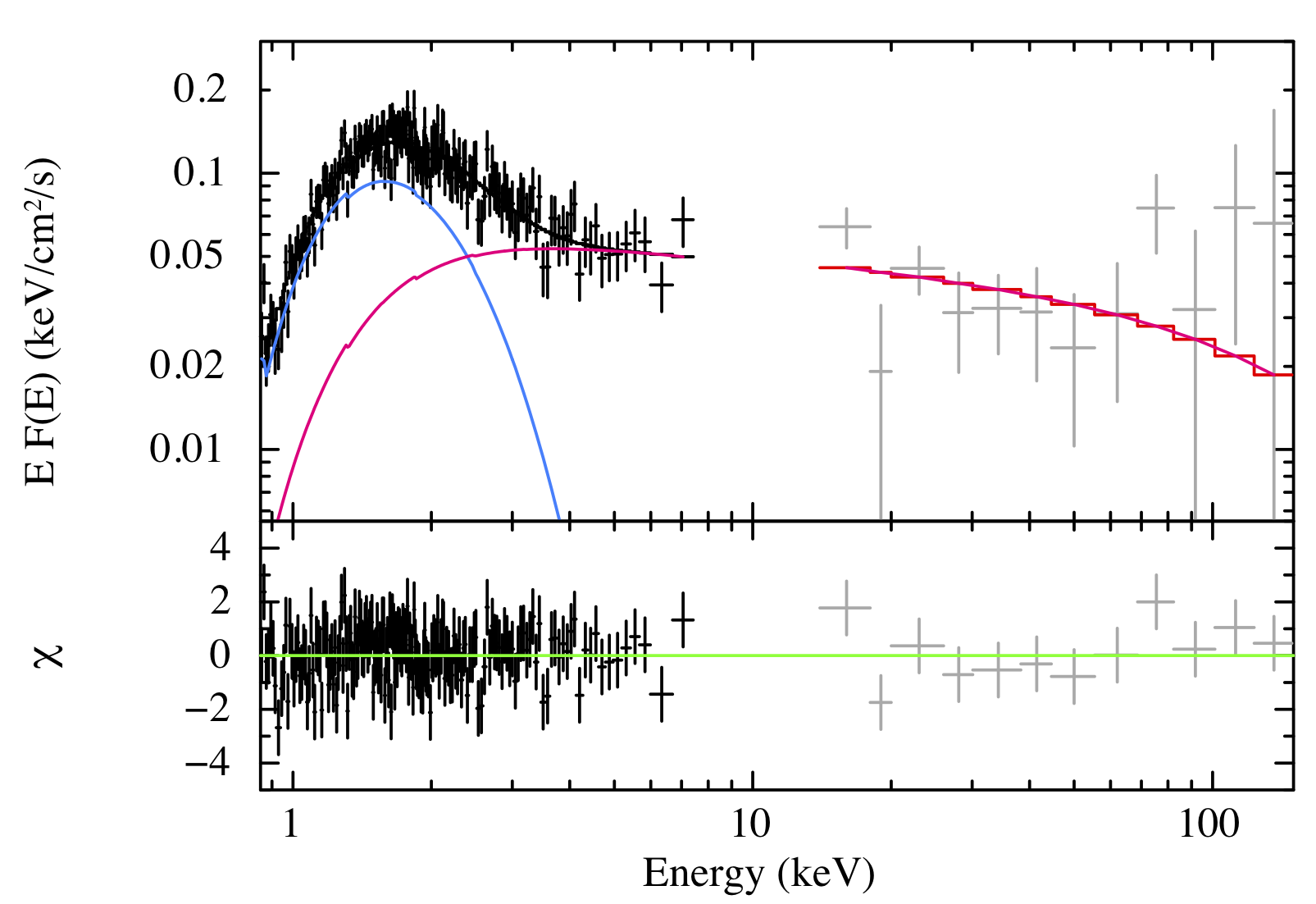}
\includegraphics[width=0.9\columnwidth]{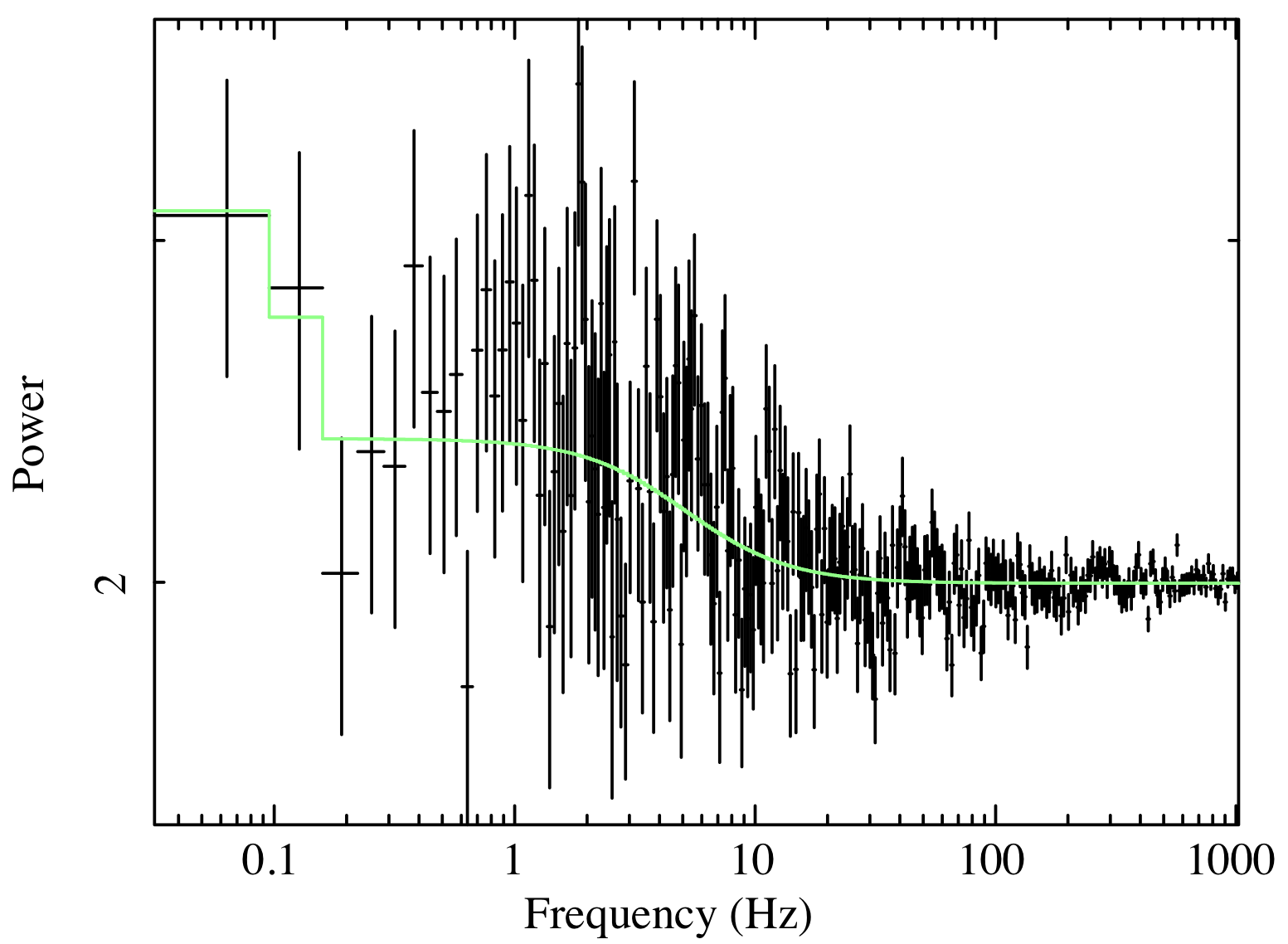}
\includegraphics[width=\columnwidth]{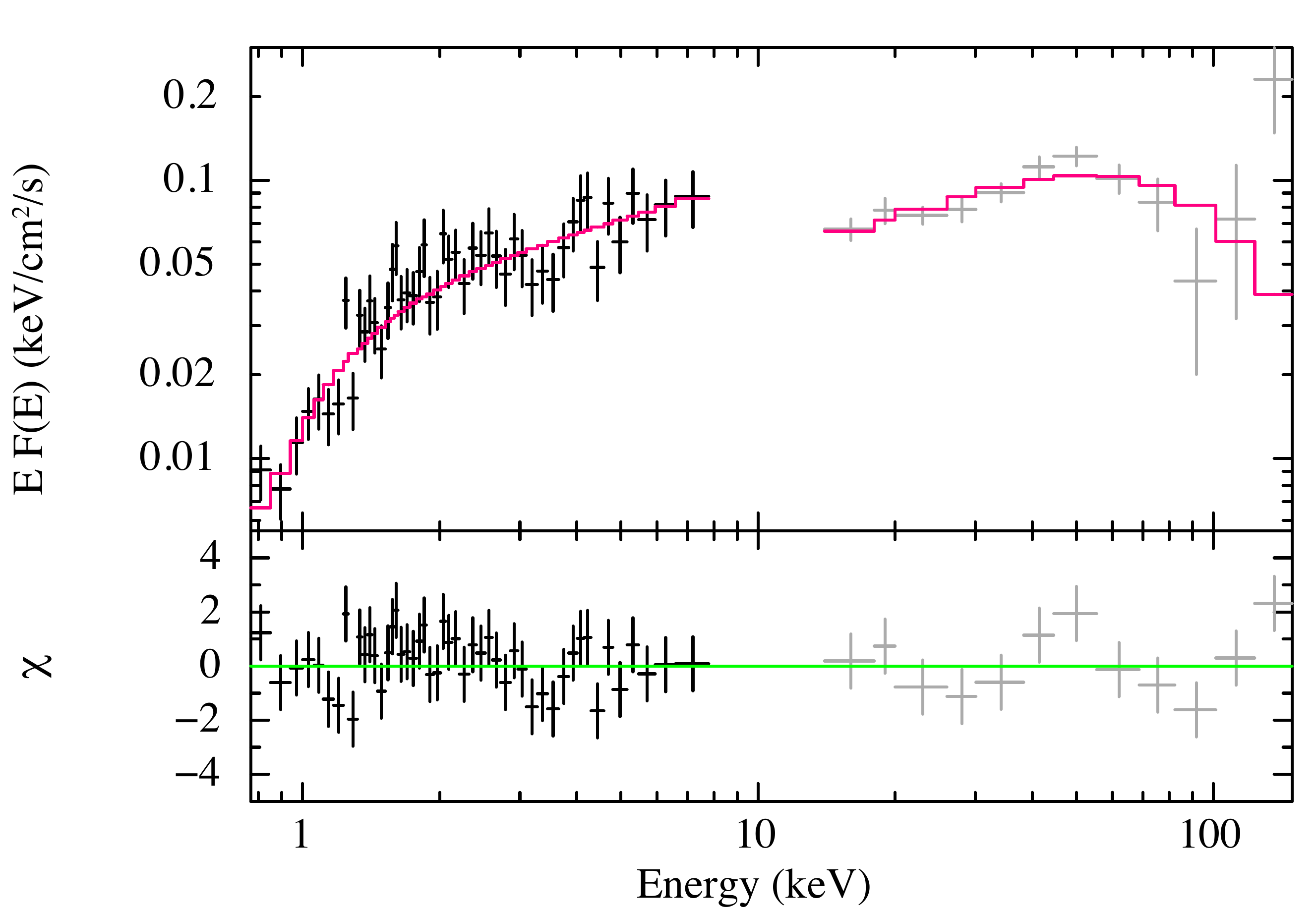}
\includegraphics[width=0.9\columnwidth]{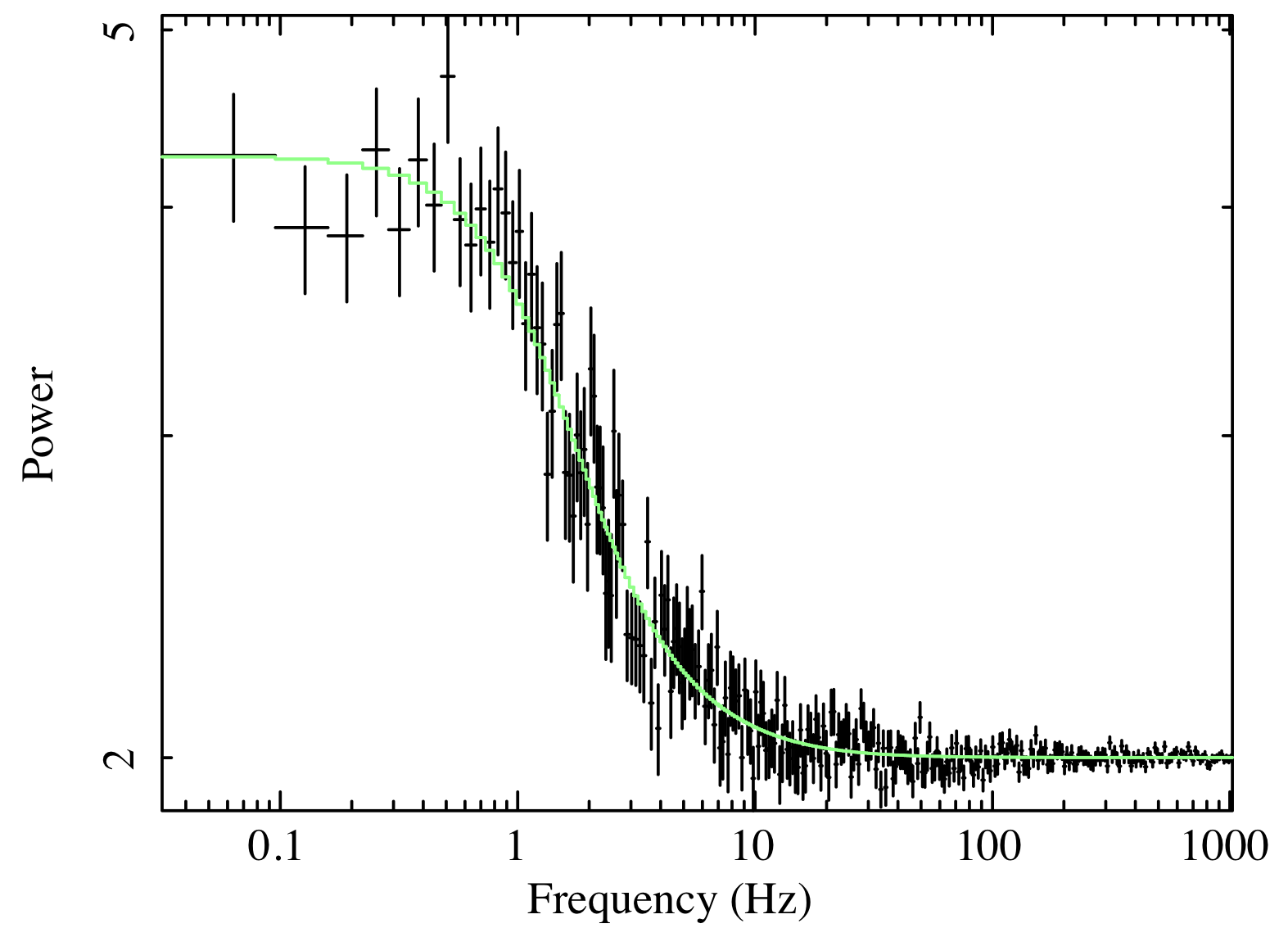}
\caption{{\textit{Left panels}:} \swift-XRT (black) and BAT (grey) spectra for a sample of three representative observations taken during each spectral state. \textit{Right panels}: Power density spectra of NICER data taken at similar times. The {\it top panels} show the X-ray observations taken around MJD~58523 (the source was not detected by \swift-BAT), during the soft state, {\it Middle panels} shows the intermediate state (around MJD~58669.76), and the {\it lower panels} are during the hard state (observations taken on $\sim$MJD~58792). Different colors show the different spectral components, where we use blue for \texttt{diskbb} and magenta for \texttt{nthcomp}. Residuals are shown at the bottom of each panel. These three representative observations show a clear change in the X-ray state.}
\label{fig:swift-spectra}
\end{figure*}

The first XRT observation occurred on 2019-02-09 (MJD~58523), showing bright X-ray emission (6$\times 10^{-10}$ erg s$^{-1}$ in the 0.5-10\,keV band; Figure~\ref{fig:lc}-\ref{fig:hid}). The spectrum of this first observation is well described with a simple \texttt{diskbb} model, with a disk temperature consistent with the \nustar\ value (Figure \ref{fig:swift-spectra}, top-left panel, and Table ~\ref{tab:broadband}). The NICER observation (obsID 1200560101), taken two days after the first XRT observation, shows an RMS consistent with zero ($<$2\%), with a PDS showing essentially only Poisson noise (see Figure~\ref{fig:swift-spectra}, top-right panel).

The 0.5 -- 10\,keV emission then faded over the next few months (see dark blue points in Figure: \ref{fig:hid}). Throughout this decay the X-ray spectra were well described with a disk blackbody model ($kT_{\rm disk}\sim$0.4-0.5 keV) and a hardness ratio of $<$0.5 (Figure~\ref{fig:lc}), implying the source being in soft state. Indeed, during this phase of the outburst, no radio counterpart to \source\ was detected during 1.28\,GHz MeerKAT observations taken on MJD~58530, with a 3$\sigma$ upper-limit of 99\,\uJybeam\ \citep{2019ATel12521....1C}.

\begin{table*}
\begin{center}
\caption{Fitting results of different spectral states. Parameters in round parentheses were kept  frozen. Quoted errors reflect 90\% confidence level. Observations are labeled for ObsID, $\dagger$=90402370, *=00011050. Parameters without a component for that fit are marked by `$-$'.}
\begin{tabular}{c c c c c c c c c c}
\hline
 Observatory & Observation & N$_H$ & $kT_{\rm in}$ & $K_{\rm disk}$ & $\Gamma$ & $kT_{\rm e}$ & \multicolumn{2}{c}{$F_{\rm unabs,0.1-100 keV}$} &$\chi^2_\nu$ \T \B\\ 
& &  ($\times$10$^{22}$ cm$^{-2}$) & (keV) & & & (keV)  &   \multicolumn{2}{c}{($\times 10^{-9}$ erg cm$^{-2}$ s$^{-1} $)} & (d.o.f.) \T \B\\
& & & & & & & \texttt{diskbb} & \texttt{nthcomp} & \\
\hline
\nustar & $\dagger$02 &  (1.0) & $0.49\pm{0.01}$ & $780^{+160}_{-130}$ & $3.45^{+0.2}_{-0.25}$ & >10 & 0.958$\pm$0.020 & 0.050$\pm0.005$ & 1.04(217) \T \B\\
XRT & *01 &  $1.34^{+0.14}_{-0.12}$ & $0.53\pm{0.03}$ & $770^{+300}_{-210}$ & - & - &  1.410$\pm{0.120}$ & - & 0.96(76) \T \B\\
XRT+BAT & *14 & $0.74\pm 0.05 $  & $0.38\pm{0.02}$ & 1800$^{+920}_{-570}$ & $2.1^{+0.3}_{-0.1}$ & >30 &  0.710$\pm$0.040 & 0.320$^{+0.060}_{-0.030}$ & 0.91(200) \T \B\\
XRT+BAT & *15 & 0.4$\pm$0.1 & - & - & $1.6 \pm 0.1$ & 24$^{+16}_{-5}$  & - & 0.960$^{+0.180}_{-0.160}$ & 1.17(55) \T \B\\
XRT+BAT & *17 & $1.0\pm0.2$ & - & -& 1.63$^{+0.07}_{-0.06}$ & >54 & - & 0.80$^{+0.10}_{-0.09}$ & 0.77(45) \T \B\\
XRT+BAT & *21 &$1.08^{+0.15}_{-0.12}$ & $0.32\pm0.02$ & 3825 $^{+4230}_{-2250}$ & 1.8$^{+0.1}_{-0.2}$ & >30 & 0.720$^{+0.080}_{-0.070}$ & 1.050$^{+0.400}_{-0.200}$ & 0.92(131)  \T \B\\
XRT+BAT & *28 & $1.13\pm0.08$ & $0.35\pm0.03$ & 3895$^{+2330}_{-1330}$ & 2.1$\pm0.3$ & (20) & 1.170$\pm$0.060 & 0.650$^{+0.040}_{-0.030}$ & 1.02(211) \T \B\\
XRT+BAT & *34 & 0.6$\pm$0.1 & - & - & $1.90^{+0.08}_{-1.0}$ & >48 & - & 0.60$\pm$0.10 & 0.71(37) \T \B\\
XRT & *41 & 1.02$\pm$0.11 & 0.55$\pm$0.03 & 660$^{+240}_{-170}$ & - & - & 1.32$^{+0.13}_{-0.12}$ & - & 0.99(72) \T \B\\ 
\hline
\end{tabular}
\label{tab:broadband}
\end{center}
\end{table*}

\subsubsection{Brightening of the hard X-ray emission}

Around MJD~58600, the 15 -- 50\,keV emission was first detected, brightening considerably over the next few months before levelling off around MJD~58750. The first XRT observation taken during this brightening in hard X-rays occurred on MJD 58670 (obsID 00011105014). We extracted a BAT spectrum averaged over an interval of $\sim$30 days centered at the time of this XRT observation, as no significant hard variability was observed in the BAT light curve during this time. We fitted these XRT and BAT spectra simultaneously. Significant emission is detected up to $\sim$100 keV (Figure~\ref{fig:swift-spectra}, middle-left panel) and is well described by thermal Comptonized emission (\texttt{nthcomp}) plus a  thermal, standard, disk emission (\texttt{diskbb}). $\Gamma = 2.1$ and a lower limit of 30 keV for the electron temperature were obtained, as the disk temperature decreased, i.e., to $kT_{\rm disk} \sim 0.4$ keV (see Table \ref{tab:broadband}). A quasi-simultaneous NICER observation performed on MJD~58672 (ID 2200560121) showed an increased RMS of $\approx$14\%, and a weak flat-top noise PDS (Figure~\ref{fig:swift-spectra}, middle-right panel), consistent with an X-ray hardening of the source. No radio observations of \source\ were taken during this phase of the outburst.

Around MJD\,58800 the hard X-ray emission was observed to briefly fade (for a few weeks). A single XRT observation showed that the soft emission had also faded around this time, when compared to the previous XRT observation. Despite the decreasing luminosity in both the hard and soft X-ray band, the broad-band X-ray spectrum was harder (Figure~\ref{fig:swift-spectra}, bottom-left panel), as shown by the increase in the X-ray hardness (Figure~\ref{fig:lc}). At this time, the disk emission becomes negligible (see Table \ref{tab:broadband}), and the Comptonisation component flattened ($\Gamma \sim$1.6), suggesting that the source was in a hard X-ray state. 
In addition, NICER observations showed the fractional X-ray emission variability progressively increased during this period, as expected for a source transitioning from the soft state, towards and then into a hard X-ray spectral state. While no X-ray quasi-periodic oscillations (QPOs) were detected from \source, NICER observations on MJD~58793 (ID 2200560140) showed a RMS of $\approx$25\% and a QPO-free flat-top noise PDS, breaking at $\approx 1$\,Hz (Figure~\ref{fig:swift-spectra}, bottom-right panel).

ATCA observations taken on MJD~58800 detected the relatively bright radio counterpart to \source\ (Table~\ref{tab:ATCA_int}). We measured radio flux densities of $570 \pm 20 \mu{\rm Jy}$ at 5.5\,GHz and $576 \pm 16 \mu{\rm Jy}$ at 9\,GHz, providing a radio spectral index, $\alpha$, of $0.1 \pm 0.2$, consistent with a flat to mildly-inverted radio spectrum (Figure~\ref{fig:lc}).

\source\ remained quite stable in the 15 -- 50\,keV X-rays band over the following $\sim$1.5 years. Towards the end of this phase, higher cadence XRT observations also show relatively stable (albeit faint) 0.5 -- 10\,keV emission, with hard X-ray spectra (X-ray hardness exceeding 1 and lack of the disk blackbody component, see Table \ref{tab:xrt_spectra}) indicating that the system was still in the hard state (see e.g. obs. *17 in Table \ref{tab:broadband}). 

\begin{figure}
\centering
\includegraphics[width=1.15\columnwidth, angle=0]{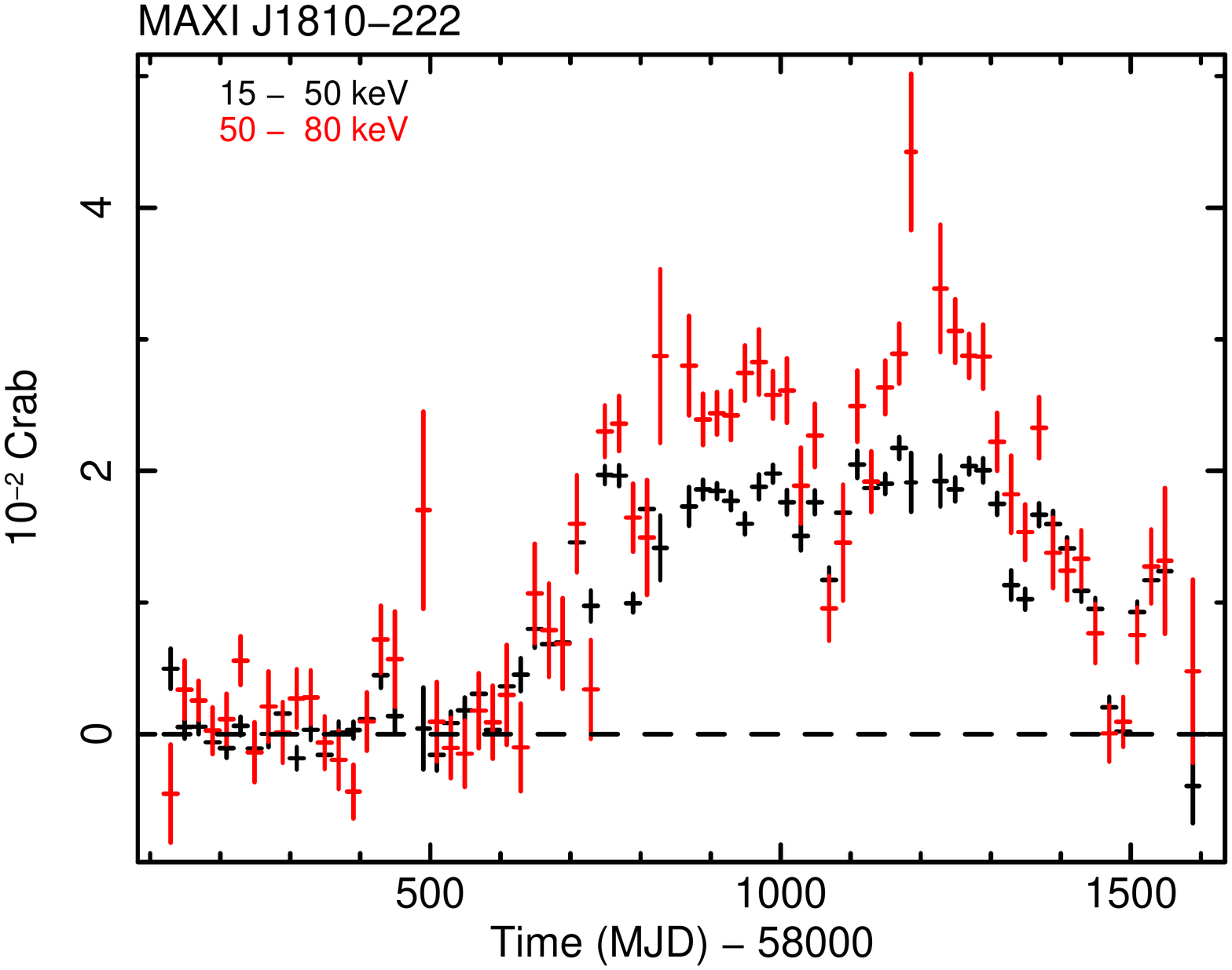}
\caption{BAT light curve showing both the 15 -- 50\,keV (black) and 50 -- 80\,keV (red) energy bands. The time binning is 20 days and the X-ray fluxes are reported in units of Crab. Bright flares were observed in the harder energy band between $\sim$MJD~58850 -- 59000 and $\sim$MJD~59150 -- 59300, which were not observed in the 15 -- 50\,keV band.}
\label{fig:bat_2bands}
\end{figure}

We also extracted the BAT lightcurve in the 50 -- 80\,keV energy band (see Figure~\ref{fig:bat_2bands}). While the 50 -- 80\,keV X-rays typically traced a similar behaviour to the 15 -- 50\,keV X-rays, we did observe some bright flares in the 50 -- 80\,keV band that were not seen in the lower X-ray energies. In particular, during the time intervals $\sim$MJD~58850 -- 59000 and $\sim$MJD~59150 -- 59300 the hardest 50 -- 80\,keV showed significant brightening above the 15 -- 50\,keV X-rays. Such flares in only the hardest X-ray bands can arise from either an increase of the electron temperature of the thermal Comptonisation, or from the appearance of an additional high energy component.

It is worth noting that the ATCA radio observations taken during the $\sim$MJD~59150 -- 59300 range showed steady, fainter radio emission, with flux densities of $\sim$60 -- 80 $\mu$Jy. Over this time the radio spectrum remained consistent with flat ($\alpha \approx 0$), although due to the faintness of the radio counterpart the errors on the radio spectrum were large.

Around MJD~59322 the hard X-ray emission once-again briefly faded, decreasing by a factor of $\sim$2 over a period of a few weeks. This time, however, it was accompanied by a brightening of the soft 0.5--10\,keV emission (which increased by a factor of $\sim$4; Figure~\ref{fig:lc}). This X-ray softening is apparent in both the X-ray hardness (Figure~\ref{fig:lc}) and the HID (Figure~\ref{fig:hid}). In the first XRT observation (obsID *21) taken during this phase, the contribution from a cold disk ($kT_{\rm disk} \sim$0.3 keV) becomes significant again and $\Gamma$ steepened to $\sim$1.8. The softer X-ray spectral behaviour exhibited during this brightening was considerably variable (see Figure \ref{fig:hid}), with the system erratically moving back and forth from high flux softer states (such as obs. *28, Table~\ref{tab:broadband}) to low flux harder states (obs. *34, Table~\ref{tab:broadband}). Only the higher X-ray flux observations require the addition of a cold disk in the X-ray spectral models (see Table \ref{tab:xrt_spectra}). As the X-rays brightened, the radio emission also increased. Throughout the brightening, the radio spectrum remained flat ($\alpha \approx 0$). Interestingly, as the source once-again faded to its pre-brightening levels (from a few weeks earlier) the radio spectrum became increasingly inverted ($\alpha = 1.3 \pm 0.7$). 

\subsubsection{Hard X-ray decline}

The hard X-rays began to fade away after $\sim$MJD~59400. Around the same time, the soft X-rays also brightened considerably, implying a return to a soft X-ray state (Figure~\ref{fig:lc} and \ref{fig:hid}). Around the time of the XRT peak (in obs. *41) the hard X-ray emission (beyond 15\,keV) became too faint to be detected. The XRT spectrum can be well described by a simple \texttt{diskbb} model, with $kT_{\rm disk}\sim$0.6\,keV, the highest disk temperature observed up to this point. The radio spectrum during the decline was measured to be steep ($\alpha = -1.1 \pm 0.5$). 

The XRT emission then faded and the spectrum hardened. This coupled with the X-ray hardness and the source evolution in the HID suggest a return to the hard state. No radio emission was detected from \source\ during this hard state. 

An additional ATCA observation taken a few weeks later showed relatively bright radio emission with a radio spectrum consistent with flat (Table~\ref{tab:ATCA_int} and Figure~\ref{fig:lc}). At the same time, BAT monitoring showed that the 15 -- 50\,keV emission had once again brightened suggesting a return to a bright hard state. However, that return was short-lived and the source faded, becoming undetected by the BAT monitoring suggesting a possible end to the outburst (at least during our monitoring). Although, due to the lack of further monitoring at soft X-rays, it is also possible that the source remained in a bright soft state beyond this point. 

\subsection{Radio position and reported possible \textit{GAIA} counterpart}\label{sec:radio}

\begin{figure}
\centering
\includegraphics[width=1\columnwidth]{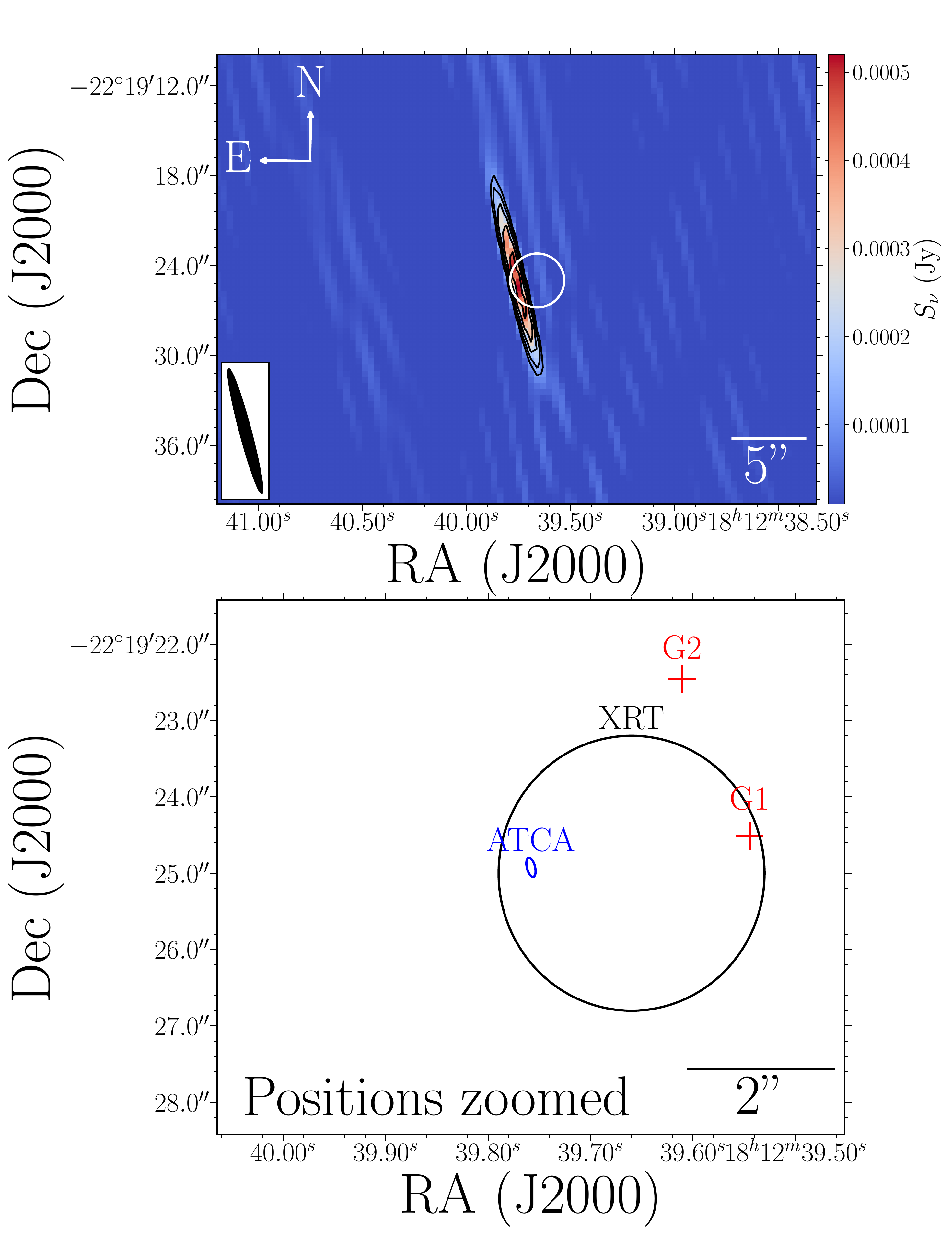}
\caption{Radio detection and source position of \source. \textit{Top panel:} The 9\,GHz ATCA image, showing the radio counterpart and the \swift-XRT position (error shown by the white circle). The black contour lines are $\sqrt{2}^n$ times the image RMS, where $n = 5, 6, 7, 8, ...$ and the RMS was 12\,\uJybeam. The ATCA beam is shown in the bottom left corner. \textit{Bottom panel:} A close-in view of the best-fit ATCA (blue ellipse) and \swift-XRT (black circle) source positions, where the errors are shown as the extent of the ellipse/circle. The red crosses (labelled G1 and G2) give the positions of the two nearest \textit{Gaia} sources, where G2 was identified as a possible counterpart to \source\ from \swift-UVOT observations \citep{2019ATel12487....1K}. The errors on the positions of G1 and G2 are on the order of milliarcseconds and are too small to be seen on this figure. Our radio position suggests that neither G1 or G2 (or any other \textit{Gaia} source) are associated with \source.}
\label{fig:position}
\end{figure}

Our measured radio position (Section~\ref{sec:radio}) is consistent with the \swift-XRT position \citep{2019ATel12487....1K}, as shown in Figure~\ref{fig:position}. From optical and UV observations, \citet{2019ATel12487....1K} proposed a possible \textit{Gaia} counterpart to \source\ (labelled as G2 in the lower panel of Figure~\ref{fig:position}), although this possible counterpart was slightly outside the XRT 90\% error region. If G2 is the optical counterpart to \source, it would imply a relatively low source distance of $730 \pm 30$\,pc. However, our radio source position is not consistent with the potential optical/UV counterpart, G2, or any other source listed in the \textit{Gaia} catalogue, where the nearest object, labelled G1, in Figure~\ref{fig:position}, lies $\approx$3$^{\prime\prime}$ from our radio position (Figure~\ref{fig:position}, lower panel)\footnote{\textit{Gaia} source IDs:\\ G1 = 4090746312693600512,\\ G2 = 4090746312736302848.}. Therefore, our radio position suggests that G2 is not associated with \source. G1, which does not have a \textit{Gaia} parallax distance, is consistent with the XRT position but is located $\approx$3.2\arcsecond\ away from our radio position. From the \textit{Gaia} (early data release 3) catalogue, the probability of finding an unrelated source within 3.2\arcsecond\ from any random position is $\approx 90\%$. Therefore, we conclude that neither G1 or G2 are associated with \source. In addition, the line of sight direction of source is towards the Galactic bulge, so it is not unusual for an optical counterpart to not be detected if the source is sufficiently distant, due to extinction close to the Galactic plane.

\section{Discussion}
\label{sec:discussion}

We have studied the broadband properties of \source. In this work, we present X-ray spectra of \source\ with \swift\ (XRT and BAT) and \nustar, as well as the X-ray timing properties during three representative spectral states with NICER. We also provide the result of our coordinated radio campaign with ATCA. While the source was initially detected in the soft state and exhibited a long and complex outburst, the source properties are consistent with an XRB passing through different X-ray spectral states \citep[see, e.g., ][]{2006csxs.book..157M,2006csxs.book..381F}. During its hard X-ray states the X-ray spectrum displayed an X-ray photon index of $\sim$1.6 and a persistent flat-spectrum radio jet. On the contrary, during the soft X-ray states the inner disk emission becomes dominant and the radio emission was either quenched or steep, as expected for a soft state \citep[e.g.,][]{2004MNRAS.355.1105F}. In addition, we have also observed a number of intermediate states with spectral parameters in between the two main states. Here we discuss a number of key properties of the system that we can determine from our multiwavelength monitoring campaign.

\subsection{Nature of the compact object}

Without an estimate on the distance to the source, identifying whether the compact object in an XRB is a BH or NS from radio and X-ray monitoring can be challenging (unless X-ray bursts or pulsations are detected). While our results from \source\ may be able to be reproduced by both a BH and NS accretor, below we argue that the X-ray behaviour of \source\ is more consistent with a BH system. 

No X-ray pulsations were detected and no trace of Type-I X-ray bursts were found in the X-ray light curves. If detected, these properties would indicate a NS accretor. However, lack of detection of these two phenomena does not necessarily mean the accretor is not a NS. A large number of the known NS XRBs do not display X-ray pulsations, possibly due to the magnetosphere being absent or very weak \citep[see, e.g. ][ and references therein]{Patruno2018} or the polar caps are not favourably aligned. Furthermore, X-ray pulsation behaviour is often intermittent \citep{2008ApJ...674L..41C,2010MNRAS.403.1426P,2012ApJ...753L..12P,2013MNRAS.432.1695C}. 
The exhibited X-ray timing and spectral properties, including the disk temperature and the soft/hard state evolution of \source\ are more suggestive of a BH primary. For NS XRBs in their soft states an additional blackbody component is generally required to account for the presence of a hot stellar surface. This additional component is typically at a higher temperature ($\sim$1--2 keV) with respect to the disk \citep[see, e.g.,][]{2001AdSpR..28..307B, Dai2010, Ludlam2018, Marino2019b}. From our monitoring, the highest temperature we derive is $\sim 0.6$ keV and an additional blackbody component was not required \citep[see, e.g.,][]{delsanto08}.

In addition, even though the estimated values of the X-ray RMS are consistent both with NS and BH systems \citep{munoz2014}, the timing parameters and PDS are more representative of an accreting BH. We do note that the lack of significant QPOs in the PDS may be surprising in the case of a BH binary. However, this could be explained in terms of a low orbital inclination of the source, which would make QPOs seen in the hard and hard intermediate states (i.e. type-C QPOs) much less prominent \citep{2015MNRAS.447.2059M}. The shape of the PDS continuum and lack of X-ray variability during the soft states are more consistent with a BH accretor \citep[see, for e.g.,][]{2019NewAR..8501524I}. Therefore, due to the spectral and timing behaviour, we favour a BH accretor in \source.

\subsection{Outburst evolution and state transitions}

In a typical outburst, BH XRBs are usually observed to brighten in a hard X-ray state, before transitioning through the hard and soft intermediate states to the soft X-ray state. This hard to soft state transition is usually observed to occur at X-ray luminosities of $\geq$3\% of the Eddington luminosity ($L_{\rm Edd}$; \citealt{2010MNRAS.403...61D}). As the outburst decays, the source then typically transitions back to the hard state at X-ray luminosities of between 0.3\% and 3\% \lEdd\ \citep{2003A&A...409..697M,2010MNRAS.403...61D,2019MNRAS.485.2744V}. However, some BH XRBs, or specific outbursts from some systems, do not follow this standard pattern of outburst: sources may not transition beyond the hard or intermediates states \citep[showing no soft states, e.g.,][]{1994ApJ...425L..17H,2004NewA....9..249B}, or can show complex outburst behaviour, such as multiple outburst peaks, re-brightenings, or glitches (e.g., \citealt{1997ApJ...491..312C,2013ApJ...775....9H,2017MNRAS.470.4298Y,2019ApJ...876....5Z, 2019ApJ...878L..28P}), and may complete the reverse transition at exceptionally low X-ray luminosities \citep{2014ApJ...791...70T,2019ApJ...883..198R,2019MNRAS.488L.129C}.

\source\ was initially detected in the soft state (e.g., \citealt{2018ATel12264....1M}), before transitioning to an apparent long-lived hard state where it remained for $\sim$1.5 years (although we do not have good coverage of the low-energy, soft X-rays over this time, which may variability indicating otherwise). After this long-lived phase, \source\ returned to the soft state, before fading at X-ray wavelengths and transitioning once-again to the hard state (Figure~\ref{fig:hid}). Afterwards, it appears that the source re-brightened again in both XRT and BAT. However, \source\ was not visible to XRT due to the position of the Sun and our broader monitoring campaign had ceased so we did not monitor the source further.

Despite the seemingly complex and unusual outburst pattern, the explanation could be simple if \source\ is located at a sufficiently large distance. In this scenario, the low/hard states typically observed at the beginning of an outburst were simply not bright enough to be detected by all-sky X-ray monitors. The  soft$\rightarrow$hard transition and subsequent hard state would then simply be a loop/excursion back to a bright hard X-ray state close to the peak of the outburst. Such loops are not unusual in BH XRBs \citep[e.g.,][]{2004MNRAS.355.1105F}, and sometimes can appear more X-ray bright than the rising hard state \citep[see][and references therein for further discussion]{2016ApJS..222...15T}. Although, we note that these loops are usually relatively short lived, lasting weeks and not years like in \source. This source has remained in outburst for a number of years, and was in a bright harder state for nearly 1.5 years (however, as mentioned previously, without soft X-ray monitoring we cannot exclude soft X-ray flaring driving state transitions). Long-duration bright hard states are not unheard of, for e.g., sources such as 4U~0540$-$697, GRS~1758$-$258, GRS~1915$+$105, 4U~1956$+$350, Swift~J1753.5$-$0127, as well as others, have been observed in outburst or within bright hard states for even longer periods of time \citep[see discussions and table 15 in][for a complete list, and references therein]{2016ApJS..222...15T}. Additionally, the recent outburst of MAXI~J1820$+$070 remained in a bright hard state for more than 100\,days during the rise phase of its outburst \citep[e.g.,][]{2021NatCo..12.1025Y}. Exploring the system properties of BH XRBs that exhibit long-lived hard states does not reveal any obvious shared properties  \citep[e.g.,][]{2016ApJS..222...15T,2016A&A...587A..61C}, with no clear connection between hard state duration and BH mass, orbital period, inclination, or mass function.

Near the end of our monitoring, \source\ appeared to follow a more standard pattern of outburst decay, where it transitioned to the soft state, began to fade, and then returned to the hard state. Our final XRT observation captured the system brightening once again, which was accompanied by a brightening in BAT. This re-brightening was short-lived however, where the BAT emission faded once again. Although it is possible the source remained bright in the soft X-ray band. Unfortunately, we were not able to follow the subsequent evolution further with XRT due to visibility and telescope constraints, and we did not observe the source again with ATCA.

\subsection{Radio emission from \source}

Our radio campaign on \source\ showed radio emission typical of a BH XRB. During the hard states the radio emission was consistent with a flat spectrum, where $\alpha \sim 0$, arising from optically-thick synchrotron emission from a self-absorbed compact jet \citep[e.g.,][]{1979ApJ...232...34B}. During the soft states the radio emission was observed to be either quenched or variable with a steep radio spectrum, indicative of an optically-thin transient jet that is launched around the hard to soft state transition \citep[e.g.,][]{2006csxs.book..381F}. 

We also note that the flat spectrum remained throughout the soft X-ray brightening ($\sim$MJD~59322), suggesting that while the X-ray spectrum did soften, the steady jet remained on, suggesting that the source likely only transitioned as far as an intermediate state.

\subsubsection{Location on the radio/X-ray plane}

\begin{figure}
\centering
\includegraphics[width=1\columnwidth]{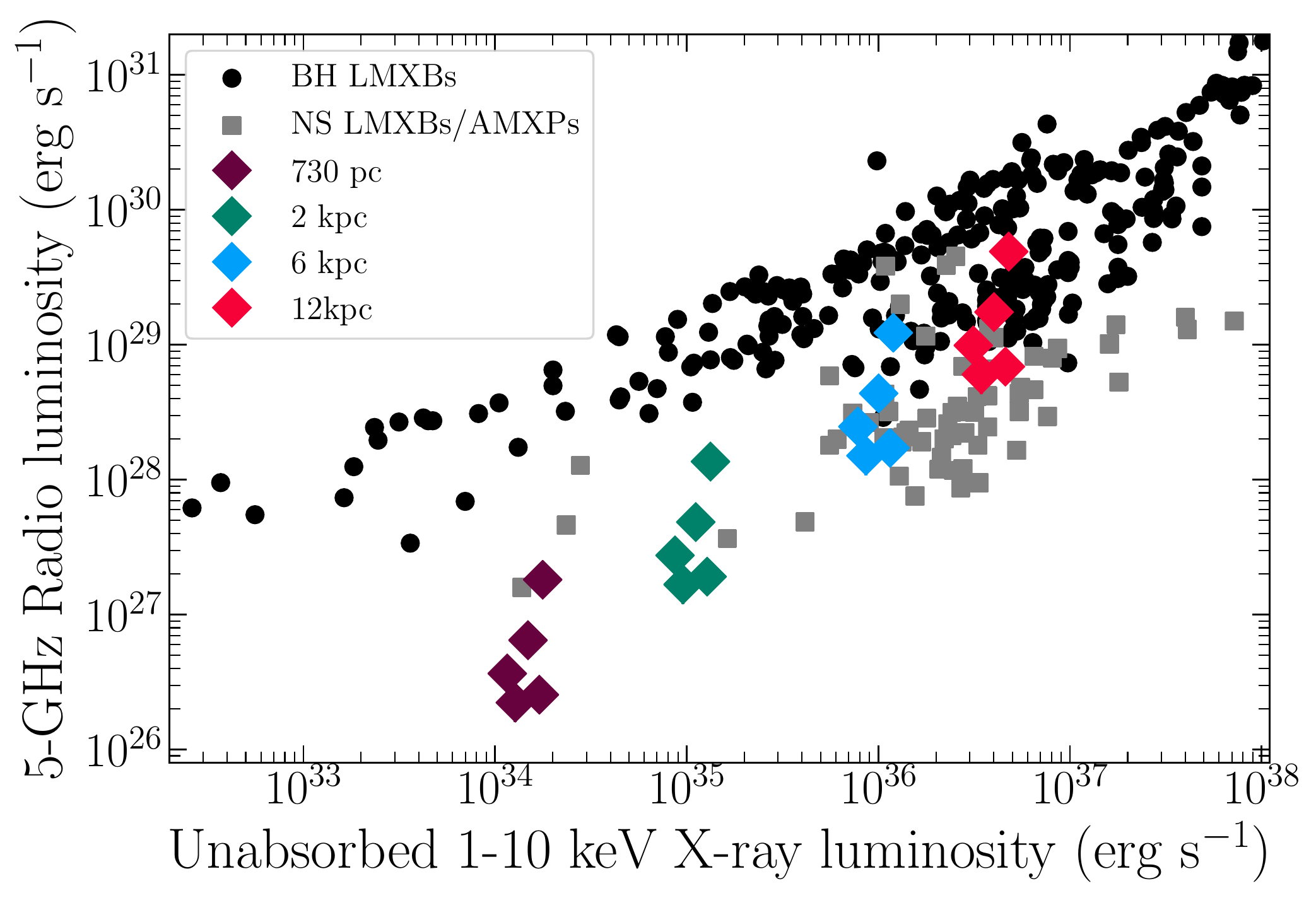}
\caption{The radio and X-ray luminosity of \source\ for varying source distances (diamonds), plotted with the full sample of accreting BHs (black circles) and NS (grey squares), where often multiple observations are plotted for a single source at different luminosities (data taken from \citealt{arash_bahramian_2018_1252036}). For completeness, we also include the sample of accreting millisecond X-ray pulsars (AMXPs) and transitional millisecond pulsars (tMSPs) within the NS sample. We only show quasi-simultaneous data taken when the source is securely in the hard X-ray state. If \source\ is indeed a BH XRB, for the observed luminosities to be consistent with other BH systems, the source would need to be located at a distance in excess of 6\,kpc.}
\label{fig:lrlx}
\end{figure}

Placing our quasi-simultaneous hard state radio and X-ray observations of \source\ on the radio/X-ray plane allows us to compare its inferred radio and X-ray luminosities against typical luminosities of BH (and NS) XRBs. Due to the unknown source distance, the observed radio and X-ray luminosities are consistent with those typically observed from both NS and BH XRBs (and we caution against using this method as the sole method to identify the nature of the compact object). However, as discussed previously, the observed X-ray spectral and timing properties suggest that \source\ is most likely a BH XRB. If that is the case, comparing its hard state radio and X-ray luminosities against those typically observed from other BH systems, to be comparable in luminosity \source\ needs to be located at a distance of $\gtrsim 6$\,kpc. We note that some systems deviate from the standard radio/X-ray luminosities shown by the broad population of BH XRBs, displaying a hybrid correlation and occupying a different parameter space (for e.g., MAXI~J1348$-$630; \citealt{2021MNRAS.505L..58C}). Therefore, we caution against its use as anything more than a suggestion that \source\ is relatively distant, although that proposition is also supported by the X-ray non-detection during the rising hard state (by \swift-BAT and MAXI). 

\begin{figure}
\centering
\includegraphics[width=1\columnwidth]{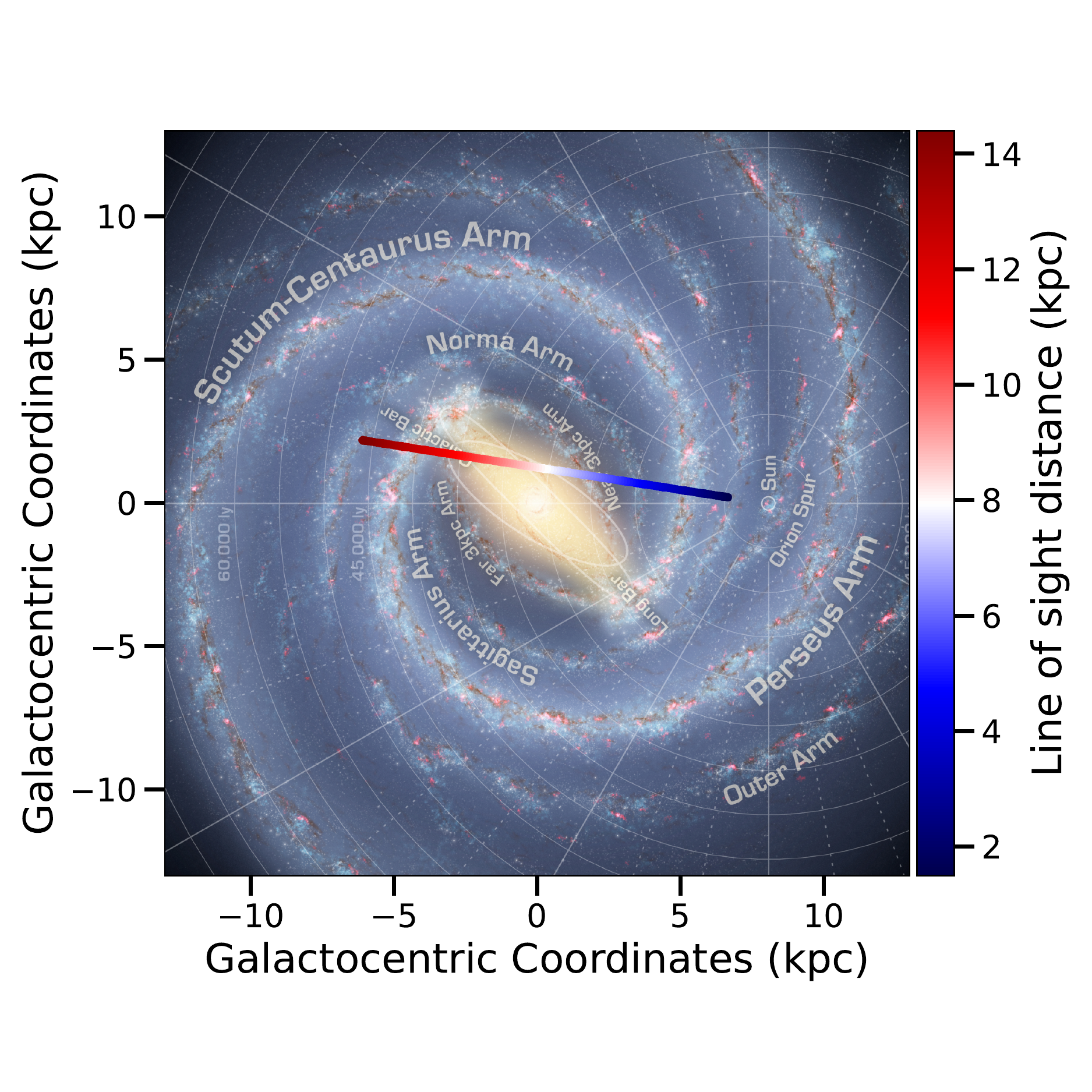}
\vspace{-8mm}
\caption{The line-of-sight towards of \source\ plotted against the Milky Way in Galactocentric coordinates. The coloured line represents the direction to \source\ for different distances to the source (as shown by the colour bar), indicating that the system can lie at large distances while remaining in the denser regions of the Galactic centre. The background image is an illustration of the Milky Way (illustration credit: NASA/JPL-Caltech/ESO/R. Hurt), image created using the python package \textsc{mw-plot}. }
\label{fig:mw_plot}
\end{figure}

Out of interest, we also show the line-of-sight direction to the source against a top-down view of the Milky Way showing that \source\ can still be located in the Galactic Bar region out to $\sim 12$\,kpc (Figure~\ref{fig:mw_plot}), implying larger distances remain plausible while still residing within higher-density Galactic regions (which does not begin until $\sim$6\,kpc in that direction). Using the Galactic mass density model of low-mass XRBs \citep{2002A&A...391..923G} as implemented in \citet{2019MNRAS.489.3116A}, we estimate the probability of the source being at distances above 6\,kpc to be 94\% (and 31\% at $>$10\,kpc). 

In support of a distance to the source in excess of 6\,kpc, our \swift-XRT gives an \nh, which traces the interstellar gas along the line of sight, of $\sim 1 \times 10^{22}$\,cm$^{-2}$. This \nh\ is more than twice the expected Galactic contribution (with a weighted average of $4.2 \times 10^{21}$\,cm$^{-2}$ being expected) along the line of sight \citep{2016A&A...594A.116H}. While a higher than expected absorption along the line of sight does not indicate the true distance to a source, it is generally thought be a result of a source's close proximity to the Galactic centre \citep[e.g.,][]{2006ApJ...646..394M}. High resolution three-dimensional optical reddening maps suggest that the majority of the Galactic extinction along the line of sight to \source\ lies between 4.5 -- 6\,kpc, suggestive that \source\ lies beyond this distance. Although, higher than expected extinction can also be a result of the source being at such a high-inclination that the disk material or donor star passes through the line of sight \citep[e.g.,][]{2002A&A...386..910P}. However, no X-ray dips are detected in the extensive X-ray monitoring of this source implying this is likely not the case. Additionally, the lack of any QPO detection is more reminiscent of a low inclination source. As such, the measured \nh\ and line-of-sight direction is suggestive that \source\ lies at larger distances, closer to the Galactic center where the higher than expected absorption may arise from within the Galactic bulge).

During its hard X-ray states, \source\ showed a large variation in radio luminosity in comparison to its X-ray evolution, where the radio luminosity was observed to change by a factor of $\sim 10$ and the X-ray only by a factor of $\sim 2$. Such large radio variation with respect to the X-rays is not regularly observed in BH systems. However, it has been observed in at least one low-inclination BH XRB, where large variations in the radio emission could be explained by variable Doppler boosting of the jet emission and a low inclination \citep{2015MNRAS.450.1745R}. As shown in Figure~\ref{fig:bat_2bands}, it is possible that the larger than expected variations in the hard state radio flux densities may be related to flaring observed in the hardest 50 -- 80\,keV X-rays, which are not captured in the standard 1--10\,keV band used for typical radio/X-ray correlations. Speculatively, there may be a suggestion of an anti-correlation between the 50 -- 80\,keV flux and the radio flux density during the bright hard state, where our brightest radio detections occurred during the fainter 50 -- 80\,keV periods. Connecting the radio flux to the hardest X-ray emission could help to understand the origin of the additional and variable soft $\gamma$-ray component observed by the {\it{INTEGRAL}} X-ray telescope in several BH XRBs during their hard states (see, e.g., \citealt{delsanto08}, \citealt{Droulans2010}). However, between $\sim$MJD~59150 and 59300 the 50 -- 80\,keV X-rays were bright and exhibiting strong variability. Over the same period, no significant radio flaring was observed, with the radio emission remaining faint and steady. If there was a simple and direct connection between the hard X-rays and the jet, we would naively expect similar changes in the radio emission. Hence, it is likely that such a connection, if there, is more complex and not observed within our data. As such, any connection between the jet and additional (to the thermal Comptonisation) hard X-ray component is speculative. A combination of multi-wavelength observations and physical models would be necessary for such a study (e.g., \citealt{Bassi2020}).

\section{Conclusions}
\label{sec:conclusions}

\source\ is a Galactic X-ray transient that was first detected in outburst in 2019. X-ray spectral and timing properties are most consistent with it being BH XRB, although a NS counterpart cannot be ruled out. At first glance, this source has displayed an odd outburst evolution, first being detected in a soft state, before transitioning to a bright hard state where it remained for $\sim$1.5 years. After which, the source returned to the soft state, faded and then transitioned to a hard state as the outburst apparently decayed. However, the seemingly odd outburst evolution can be simply explained with a sufficiently large distance to the source, such that the source was not detectable by all sky X-ray monitors (which primarily work in the harder X-ray bands) during its lower-luminosity rise phase. In this scenario, the initial bright soft$\rightarrow$hard state transition could be a result of a loop/excursion back to the hard state from the high soft state, and not a standard reverse transition during the outburst decay. Such excursions are not rare in BH systems. At later times, the source appeared to follow a more standard X-ray evolution. Our radio detection also rules out the proposed association with a proposed \textit{Gaia} counterpart \citep{2019ATel12487....1K}.

\section*{Acknowledgements}

We would like to thank the anonymous referee for their helpful comments. Authors @INAF acknowledge financial contribution from the agreement ASI-INAF n.2017-14-H.0 and INAF mainstream. This work has been partially supported by the ASI-INAF program I/004/11/5. The Australia Telescope Compact Array is part of the Australia Telescope National Facility which is funded by the Australian Government for operation as a National Facility managed by CSIRO. We acknowledge the Gomeroi people as the traditional owners of the Observatory site. This research has made use of (i) NASA's Astrophysics Data System, (ii) data, software, and/or web tools obtained from the High Energy Astrophysics Science Archive Research Center (HEASARC), a service of the Astrophysics Science Division at NASA Goddard Space Flight Center (GSFC) and of the Smithsonian Astrophysical Observatory's High Energy Astrophysics Division, (iii) data supplied by the UK \textit{Swift} Science Data Centre at the University of Leicester, and (iiii) MAXI data provided by RIKEN, JAXA and the MAXI team. This work has made use of data from the European Space Agency (ESA) mission {\it Gaia} (\url{https://www.cosmos.esa.int/gaia}), processed by the {\it Gaia} Data Processing and Analysis Consortium (DPAC,
\url{https://www.cosmos.esa.int/web/gaia/dpac/consortium}). Funding for the DPAC has been provided by national institutions, in particular the institutions participating in the {\it Gaia} Multilateral Agreement. AM is supported by the H2020 ERC Consolidator Grant “MAGNESIA” under grant agreement No. 817661 (PI: Rea) and National Spanish grant PGC2018-095512-BI00. This work was also partially supported by the program Unidad de Excelencia Maria de Maeztu CEX2020-001058-M, and by the PHAROS COST Action (No. CA16214).

\section*{Data Availability}

Data from \swift\ are publicly available from HEASARC (\url{https://heasarc.gsfc.nasa.gov/}). Best-fit X-ray parameters are provided in the manuscript. Raw ATCA data are provided on the Australia Telescope Online Archive (\url{https://atoa.atnf.csiro.au/query.jsp}), under project code CX445. All calibrated radio flux densities used in this work are provided.



\bibliographystyle{mnras}
\bibliography{bib} 

\bsp	
\label{lastpage}
\end{document}